\documentclass[reprint,pra, superscriptaddress]{revtex4-1}
\usepackage{amsmath,braket,subfigure,svg}
\usepackage{graphicx}
\usepackage{amsmath}
\usepackage{amsfonts}
\usepackage{natbib}
\usepackage{siunitx}
\usepackage{textcomp}
\usepackage{bm}
\usepackage{color}
\usepackage[hidelinks,colorlinks=true,urlcolor=blue,linkcolor=blue,citecolor=blue]{hyperref}
\usepackage{float}
\floatstyle{plaintop}
\restylefloat{table}

\def\ba{{\boldsymbol a}}
\def\brho{{\boldsymbol \brho}}
\def\bk{{\boldsymbol k}}

\def\br{\boldsymbol{r}}

\def\bu{{\boldsymbol u}}
\def\bv{{\boldsymbol v}}
\def\bPhi{{\boldsymbol \Phi}}
\def\bphi{{\boldsymbol \phi}}
\def\bpsi{{\boldsymbol \psi}}
\def\bJ{{\boldsymbol J}}

\def\bV{{\boldsymbol V}}
\def\bfu{\bm {\mathfrak {u}}}
\def\la{\langle}

\def\calK{\mathcal{K}}

\def\calH{\mathcal{H}}
\def\calC{\mathcal{C}}
\def\calD{\mathcal{D}}

\def\calA{\mathcal{A}}
\def\calE{\mathcal{E}}
\def\calJ{\mathcal{J}}

\def\pa{\partial}
\def\nn{\nonumber}

\def\la{\langle}
\def\ra{\rangle}
\begin{document}
\title{Intrinsic anomalous Hall effect across the magnetic phase transition of a spin-orbit-coupled Bose-Einstein condensate}
\author{Canhao Chen}
\altaffiliation{These authors contributed equally to this work.}
\affiliation{Shenzhen Institute for Quantum Science and Engineering, Southern University of Science and Technology, Shenzhen 518055, China.}
\affiliation{International Quantum Academy, Shenzhen 518048, China.}
\affiliation{Guangdong Provincial Key Laboratory of Quantum Science and Engineering, Southern University of Science and Technology, Shenzhen 518055, China.}
\author{Guan-Hua Huang}
\altaffiliation{These authors contributed equally to this work.}
\affiliation{Department of Physics, Southern University of Science and Technology, Shenzhen, 518055, China}
\affiliation{Shenzhen Institute for Quantum Science and Engineering, Southern University of Science and Technology, Shenzhen 518055, China.}
\author{Zhigang Wu}
\email{wuzg@sustech.edu.cn}
\affiliation{Shenzhen Institute for Quantum Science and Engineering, Southern University of Science and Technology, Shenzhen 518055, China.}
\affiliation{International Quantum Academy, Shenzhen 518048, China.}
\affiliation{Guangdong Provincial Key Laboratory of Quantum Science and Engineering, Southern University of Science and Technology, Shenzhen 518055, China.}

\date{\today }
\begin{abstract}
We study theoretically the zero temperature intrinsic anomalous Hall effect in an experimentally realized 2D spin-orbit coupled Bose gas. For anisotropic atomic interactions and as the spin-orbit coupling strength increases, the system undergoes a ground state phase transition from states exhibiting a total in-plane magnetization to those with a perpendicular magnetization along the $z$ direction. We show that finite frequency, or ac,  Hall responses exist in both phases in the absence of an artificial magnetic field, as a result of finite inter-band transitions.  However, the characteristics of the anomalous Hall responses are drastically different in these two phases because of  the different symmetries preserved by the corresponding ground states. In particular, we find a finite dc Hall conductivity in one phase but not the other. The underlying physical reasons for this are analyzed further by exploring relations of the dc  Hall conductivity to the system's chirality and Berry curvatures of the Bloch bands. Finally, we discuss an experimental method of probing the anomalous Hall effect in trapped systems. 
\end{abstract}
\maketitle
\section{Introduction}
The attempts to understand the origins of various non-ordinary Hall effects have played an important role in establishing the field of electronic topological materials~\cite{Hasan2010,Qi2011,2013Bernevig}. Indeed, the famous work by Thouless {\it et al.}~\cite{1982TKNN}, which was the first to introduce the concept of topological invariant for magnetic Bloch bands, originated from an analysis of the integer quantum Hall effect. Another phenomenon that has generated great interest in the context of topological materials is the anomalous Hall effect (AHE)~\cite{2010Nagaosa,2010Xiao}, which refers to strong Hall responses in the absence of a magnetic field and was originally discovered in ferromagnetic materials with spin-orbit (SO) couplings.  Central to the study of both effects is the idea that the Bloch bands occupied by the electrons have nontrivial geometric and topological features which can significantly affect transport properties such as the Hall conductivity~\cite{2010Xiao}. 

In recent years, much progress has been made in the area of quantum gases to generate such band structures~\cite{Cooper2019}, by introducing artificial magnetic fields, engineering synthetic spin-orbit couplings or some other means.  In addition to simulations of non-interacting fermionic topological phases~\cite{Aidelsburger2013,Miyake2013,atala2013,Jotzu2014}, this has enabled creations of bosonic topological systems where atomic interactions are essential for the emergence of topological properties~\cite{Xu2016, 2016Kock, 2016XiaopengLi, 2016Liberto,2007Muller,2010Wirth,2012Soltan,2013Olschlager, 2015Kock,2016Sengstock,Sun2018,2021Hachmann,2021Vargas,2021Jin,2021Wang}. In parallel, many probing methods have also been devised and implemented to measure transport properties of the quantum gas systems~\cite{Chien2015,2015Wu,2019Anderson}. These developments have led us to ask whether nontrivial Hall responses such as the AHE can be found in these bosonic topological systems and, if so, how can they be detected in experiments. Indeed, some of us have recently predicted~\cite{Huang2022} that a ground state intrinsic AHE exists in a bosonic chiral superfluid in a boron nitride optical lattice~\cite{2021Wang} and may be probed with currently available experimental techniques. We have also shown that the superfluid's chirality and the Berry curvature of the condensate mode are two factors underlying the intrinsic AHE~\cite{Huang2022}.

The main purpose of this paper is to apply the ideas and methods developed in Ref.~\cite{Huang2022} to investigate another recently realized bosonic topological system, a 2D spin-orbit-coupled (SOC) Bose gas in a optical lattice~\cite{Sun2018}. As we are interested in the AHE, we focus on the situation where the Zeeman field is absent. In this case, it was shown earlier that the interplay of atomic interactions and the SOC can give rise to nontrivial excitation band topology as well as gapless edge states~\cite{Huang2021}, both of which are also found in the bosonic chiral superfluid~\cite{2021Wang}. More importantly, a careful analysis of the ground states of the SOC system shows that it carries a finite, global angular momentum in one of the phases, making it also a chiral superfluid similar to that in Ref.~\cite{2021Wang}.  Motivated by these similarities, we perform a systematic calculation on the ground state Hall conductivity of the SOC system and reveal surprisingly even richer phenomenology of the AHE compared to the chiral superfluid studied in Ref.~\cite{Huang2022}. 

The rest of the paper is organized as follows. In Sec.~\ref{system} we provide a detailed discussion of the system's Hamiltonian and the various symmetries exhibited by the Hamiltonian. In Sec.~\ref{PT}, we calculate the ground state of the system and demonstrate the existence of a magnetic phase transition driven by the spin orbit coupling strength. The two magnetic phases are distinguished not only by different directions of the magnetization but more importantly by  different symmetries preserved by the ground state. The consequence of these distinctions are explored in Sec.~\ref{FAHE} in the context of the anomalous Hall effect, where we find contrasting structures of the ac Hall conductivity in these two phases. In particular, the dc Hall conductivity is finite in the perpendicular magnetization phase but vanishes in the in-plane magnetization phase. The underlying causes for this are analyzed in Sec.~\ref{DCHALL}. In Sec.~\ref{EP}, we discuss an experimental proposal to observe the intrinsic AHE in the SOC system using the center of mass oscillations. The main results are summarized again in Sec.~\ref{CR}. 
\section{Qausi-2D Spin-orbit-coupled Bose gas}
\label{system}
\subsection{Hamiltonian}
\begin{figure}[htbp]
	\centering
	\includegraphics[width=8.6cm]{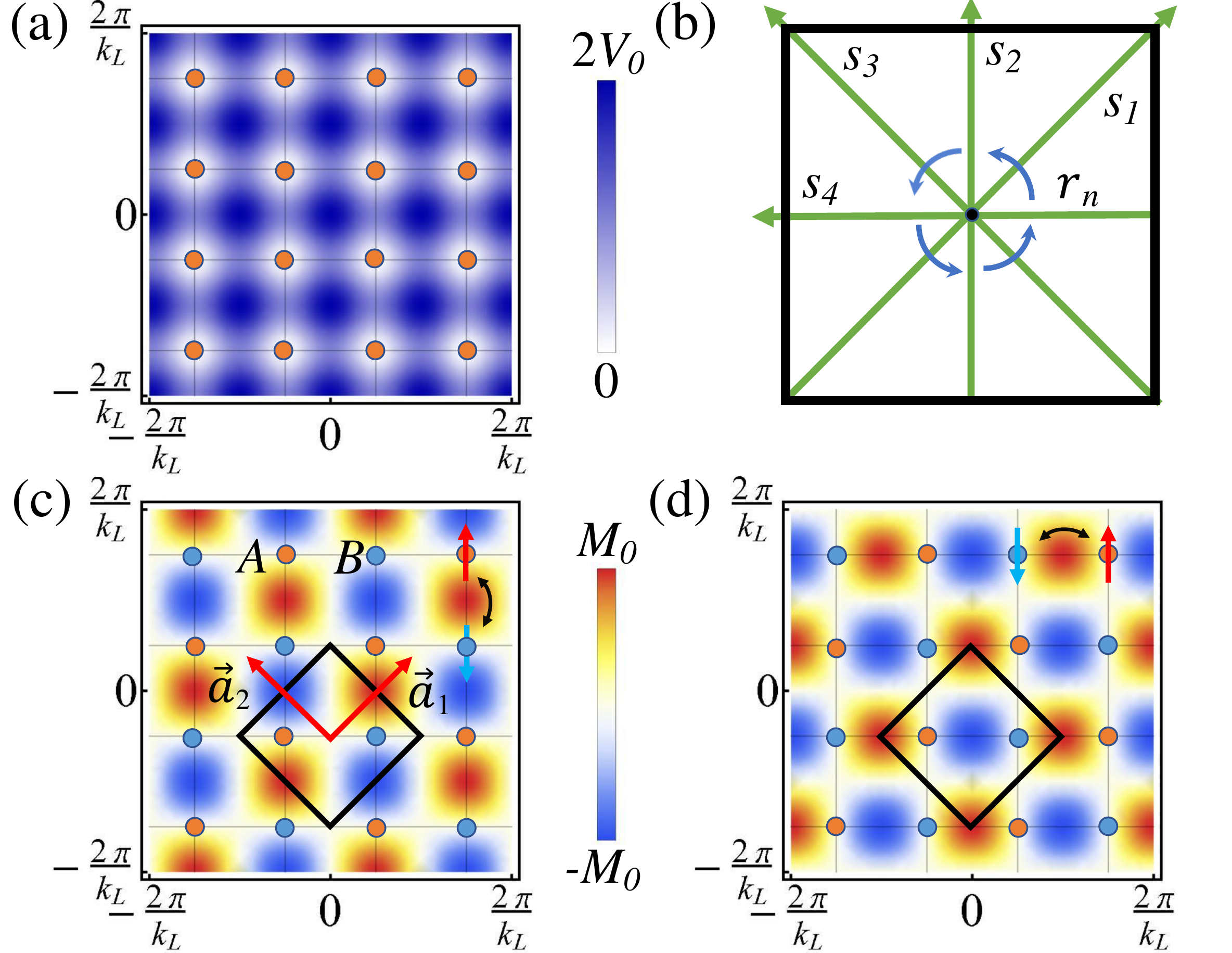}
	\caption{(a) Illustration of the optical square lattice potential.  (b) The dihedral point group symmetry $D_4$ for the single particle Hamiltonian in the absence of the Raman potentials. (c) and (d)  Illustrations of the Raman potentials $V_{R,1} (\br)$ and  $V_{R,2} (\br)$ respectively.  The Wigner-Seitz cell and primitive vectors of the Raman potentials are also shown.} 
	\label{fig1}
\end{figure}

The system of interest is a quasi-2D two-component  Bose gas confined in a square optical lattice potential, where a robust spin-orbit coupling is realized by means of a Raman lattice scheme~\cite{Sun2018}. The non-interacting part of the Hamiltonian is 
\begin{align}
\label{H_0}
\hat H_0 =& \int d\bm r\, \hat{\bm\psi}^{\dagger}(\bm r)h_{0}\hat {\bm\psi }(\bm r).
\end{align}
Here we adopt a spinor notation $\hat \bpsi =  (\hat\psi_\uparrow,\hat \psi_\downarrow)^T$, where $\hat\psi_{\sigma}$ is the field operator with  $\sigma=\uparrow,\downarrow$ as the spin index. The single-particle Hamiltonian is given by
\begin{equation}
	\label{h0}
	h_0=\left[\frac{\bm p^2}{2m}  +V_{\text{latt}}(\br) \right ] I +V_{R,1}(\br)\sigma_x + V_{R,2}(\br)\sigma_y,
\end{equation}
where $I$ is the unit matrix in the spin space and  $\sigma_{x},\,\sigma_{y}$  are Pauli matrices. Here  $V_{\text{latt}}(\br)=V_0(\cos^2k_{L}x+\cos^2k_{L}y)$ is the 2D square optical potential with lattice depth $V_0$ and lattice spacing $\pi/k_L$ (see Fig.~\ref{fig1}(a));  $ V_{R,1}(\br)=M_0\sin (k_{L}x)\cos (k_{L}y)$ and $V_{R,2}(\br) = M_0\sin (k_Ly)\cos (k_{L}x)$ are the Raman potentials coupling the spin to the motional degrees of freedom (see Fig.~\ref{fig1}(c) and (d)), where the coupling strength $M_0$ can be experimentally tuned. The Raman lattices are commensurate with the square optical lattice but have a unit cell twice as large. The primitive vectors of the composite lattice potential are thus given by those of the Raman lattices, i.e., $\ba_1 =\pi /k_L (1,1)$ and $\ba_2 =\pi /k_L (-1,1) $.  The original square lattice can then be divided into two sublattices, $A$ and $B$, distinguished by the local Raman potentials around the lattice sites. Note that we have not included any Zeeman field in Eq.~(\ref{h0}) in order to explore effects in which the atomic interactions play a fundamental role.  The latter are described by 
\begin{align}
\hat H_{\rm int} =\frac{1}{2}\sum_{\sigma\sigma'}g_{\sigma\sigma'}\int d\bm r\, \hat\psi^{\dagger}_{\sigma}(\bm r)\hat\psi^{\dagger}_{\sigma'}(\bm r)\hat\psi_{\sigma'}(\bm r)\hat\psi_{\sigma}(\bm r),
\label{Hint}
\end{align}
where $g_{\sigma\sigma'}$ are species-dependent interaction strengths. We shall consider the case of anisotropic interactions found in experiments, more specifically the case where the intra-species interaction strengths $g_{\uparrow\uparrow}= g_{\downarrow\downarrow}$ are greater than the inter-species ones $g_{\uparrow\downarrow}= g_{\downarrow\uparrow}$. As we shall see later, for such an interaction the system experiences a ground state magnetic phase transition as the spin-orbit coupling strength increases.  For the purpose of symmetry analysis, it is useful to rewrite Eq.~(\ref{Hint}) as 
\begin{align}
\hat H_{\rm int} =\frac{1}{2}\int d\bm r\, : \left [\bar g (\hat\bpsi^\dag \hat\bpsi)^2 +\delta g (\hat\bpsi^\dag\sigma_z\hat\bpsi)^2 \right ] :,
\label{Hint2}
\end{align}
where $\bar g \equiv (g_{\uparrow\uparrow} +g_{\uparrow\downarrow} )/2$, $\delta g \equiv  (g_{\uparrow\uparrow} - g_{\uparrow\downarrow} )/2$ and $:\cdots :$ denotes normal order of the field operators. 

Lastly, we mention that the total Hamiltonian $\hat H = \hat H_0 + \hat H_{\rm int}$ can be mapped to one that has the periodicity of the optical lattice potential by means of a unitary transformation~\cite{Pan2016} 
\begin{align}
 U \hat \bpsi (\br)U^{-1}= \begin{pmatrix} 
1 & 0 \\
 0 & e^{-ik_Lx - i k_L y}  
 \end{pmatrix}\begin{pmatrix} 
\hat \psi_\uparrow(\br) \\
\hat \psi_\downarrow(\br) 
 \end{pmatrix}. 
\end{align}
However, we will not adopt such a transformation here. 
\subsection{Symmetry}
A through analysis of the symmetries exhibited by the Hamiltonian is crucial to the understanding of the magnetic phase transition, since different phases can be distinguished by, in addition to the order parameter, the specific symmetries that are spontaneously broken. Furthermore, knowledge of these symmetries is rather useful in the calculation of various dynamical response functions of the system, as they ultimately determine the selection rules obeyed by the relevant dynamical transitions between the states. Pertinent to our  study are the following three types of symmetries (or symmetry groups): the Pseudo $\mathcal {PT}$ symmetry, the modified dihedral $\tilde D_4$ symmetry group and the nonsymmorphic symmetry. In the following, we describe each of these symmetries separately.

{\bf {Pseudo $\mathcal {PT}$ symmetry}}:   For genuine spin $1/2$ particles, the $\mathcal {PT}$ symmetry operation is defined as 
\begin{align}
\mathcal{PT}=i\sigma_{y}\mathcal{KI},
\end{align} where $\mathcal I$ is the space inversion operator and $\mathcal K$ is the complex conjugation operator. Because the spin in our system is pseudo-spin, we refer to this as the pseudo $\mathcal {PT}$ symmetry. It is straightforward to show that the single particle Hamiltonian $h_0$ in Eq.~(\ref{h0}) is indeed invariant under such a transformation, i.e., 
$
(\mathcal {PT} )h_0 (\mathcal {PT})^{-1} = h_0$, which ensures that $\mathcal {PT}$ commutes with $\hat H_0$ in Eq.~(\ref{H_0}).  In addition, it is clear that $\mathcal {PT}$ also commutes with $\hat H_{\rm int} $ in Eq.~(\ref{Hint2}) due to the fact that $(\mathcal {PT} )\sigma_z(\mathcal {PT})^{-1} = -\sigma_z$.

{\bf {Modified dihedral symmetry}}:  In the absence of the SO coupling, $h_0$ has the usual point group $D_4$ symmetry, which consists of four-fold rotation operations $r_n$ and  two-fold reflection operations $s_n$  ($n = 1,\cdots,4$) as illustrated in Fig.~\ref{fig1}(b). Here $r_n$ denotes the counterclockwise rotation of $n\pi/2$ around the $z$-axis at the origin and $s_n$ denotes the reflection across a line that makes an angle of $n\pi/4$ with the $x$-axis.  With SO coupling, however, the Hamiltonian $h_0$ no longer commutes with these operations because the Raman potentials are not fully invariant under $D_4$ operations. Instead, as can be checked easily, it commutes with what we will refer to as the modified dihedral symmetry operations
\begin{align}
\tilde{r}_n\equiv  e^{-i\frac{n\pi}{4}\sigma_z}r_n,\qquad \tilde{s}_n\equiv  e^{-i\frac{\pi}{2}\vec s_n \cdot \vec \sigma}s_n
\end{align}
for $n = 1,\cdots,4$ and 
\begin{align}
\tilde{r}_n\equiv  e^{-i\frac{n\pi}{4}\sigma_z}r_{n-4},\qquad \tilde{s}_n\equiv  e^{i\frac{\pi}{2}\vec s_{n-4} \cdot \vec \sigma}s_{n-4}
\end{align}         
for $n = 5,\cdots,8$, where $\vec s_{n}$ is the unit vector along the reflection axis of the $s_n$ operation. These $16$ operations form a symmetry group of $h_0$ denoted by $\tilde D_4$, which is  a double group of $D_4$ point group~\cite{Group_Cornwell,Dresselhaus2007group}. Again, since each of the operators in $\tilde D_4$ either commutes or anti-commutes with $\sigma_z$, the whole $\tilde D_4$ group commutes with $\hat H_{\rm int}$. 

{\bf {Nonsymmorphic symmetry}}: Certain crystal structures are invariant under a combination of point group rotation and non-primitive lattice translation, which is known as the nonsymmorphic symmetry.  An analogous symmetry exists for our system, where the symmetry operations are described by
\begin{align}
 \Lambda_i = T_i({\pi}/{k_L})e^{-i\frac{\pi}{2}\sigma_{z}},
 \end{align} 
where $T_{i}(l)$ $(i = x,y)$ is a translation along the $i$-direction of a distance $l$. It can be again checked that such operations commute with $\hat H_0$ and $\hat H_{\rm int}$. In fact, any combination of $\Lambda_i$ and a rotation operator in $\tilde D_4$ group, i.e., 
\begin{align}
\Lambda_i R \qquad \forall R \in \tilde D_4,
\end{align} is also a nonsymmorphic symmetry operation.

\section{Magnetic phase transition: in-plane vs. perpendicular magnetization}
\label{PT}
As the SO coupling strength $M_0$ increases, the condensate in the optical lattice undergoes a phase transition~\cite{Sun2018} which is analogous to the stripe-to-plane-wave phase transition found in translationally invariant SOC systems~\cite{Zhai_2015}. However, since the atoms in our system do not condense at finite crystal momenta nor do they form density waves, the nomenclature `plane wave phase'  and `stripe phase' are not entirely appropriate here~\footnote{This phase transition can be viewed as a stripe-to-plane-wave phase transition with respect to the transformed Hamiltonian $\hat H' = U\hat H U^{-1}$.}. Instead, we shall see that the phases are characterized by different types of the magnetization, i.e., in-plane vs. perpendicular magnetization. 
In this section we provide a detailed discussion of this phase transition and reveal important, previously unnoticed  properties of these phases which, upon comparison to those of a recently studied chiral superfluid, suggest that the anomalous Hall effect may  exist in this system. 

\begin{figure}[tbp]
	\centering
	\includegraphics[width=8.4cm]{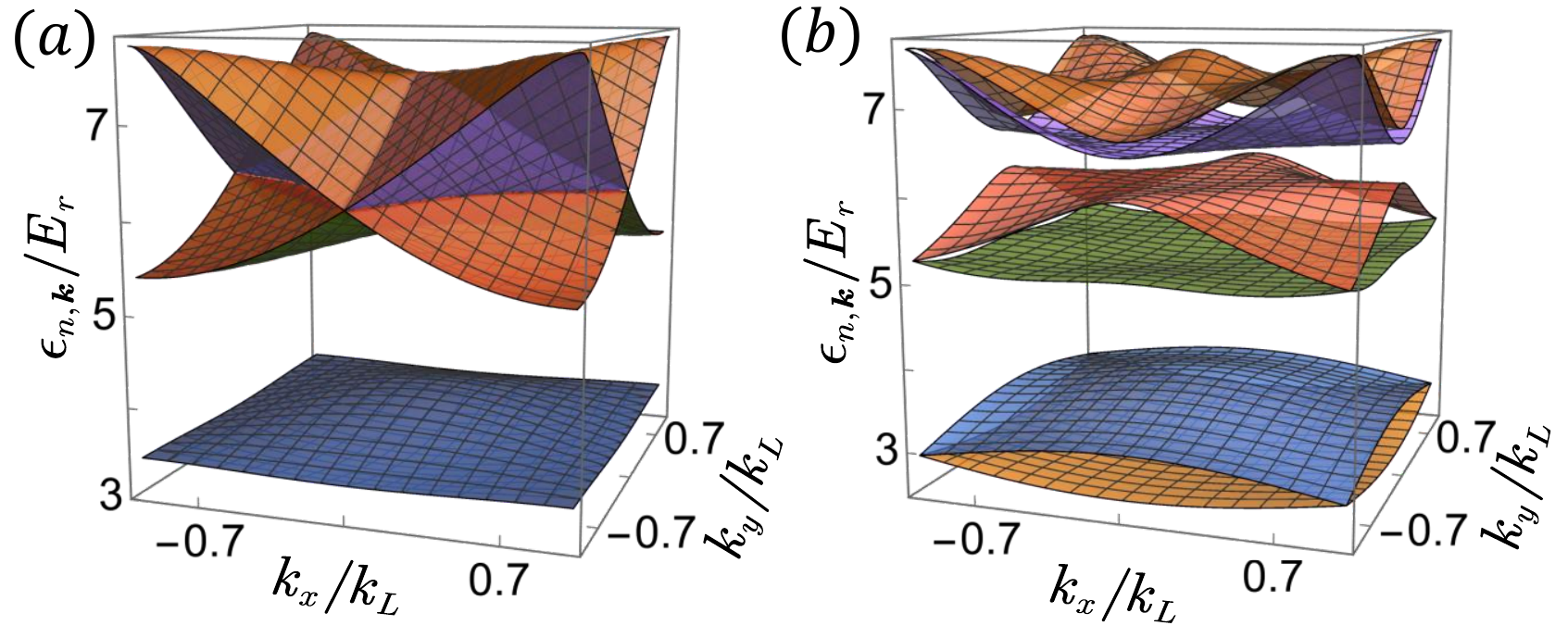}
	\caption{The first $12$ energy bands of the single particle Hamiltonian: (a) without SO coupling (b) with SO coupling. Here $V_0 = 4 E_r$ and $M_0 = 2$, where $E_r$ is the recoil energy. Note that the folded structure of the energy bands for the Hamiltonian without the SO coupling is due to the fact that it is plotted within the first Brillouin zone of the SOC Hamiltonian, which is half the size of that without SO coupling.} 
	\label{fig2}
\end{figure}

To begin, we first calculate the non-interacting bands of the SOC gas determined by 
 \begin{align}
 h_0{\bm\phi}_{n\bk}(\br) = \epsilon_{n\bk}{\bm\phi}_{n\bk}(\br),
 \label{spw}
 \end{align} 
 where $\epsilon_{n\bk}$, measured in recoil energy $E_r = {\hbar^2 k_L^2}/{2m}$, is the band dispersion and ${\bm\phi}_{n\bk}(\br)= \left [\phi_{n\bk\uparrow}(\br),\phi_{n\bk\downarrow} (\br )\right ]^T$ is the corresponding two-component Bloch state. Shown in Fig.~\ref{fig2}(b) is an example of the band structure obtained from solving Eq.~(\ref{spw}); as a comparison, the band structure in the absence of the SO coupling is shown in Fig.~\ref{fig2}(a). An important property of the non-interacting SOC bands is that they are doubly degenerate due to the $\mathcal{PT} $ symmetry, much like the familiar Kramer's degeneracy. This is because  $(\mathcal {PT})^2=-1$ and as a result $\mathcal{PT}  {\bm\phi}_{n,\bk} \neq  {\bm\phi}_{n,\bk}$.  The degeneracy is reflected in Fig.~\ref{fig2}(b) which contains the lowest $12$ bands. In practice, it's necessary to work with a specific set of single-particle wavefunctions. This can be obtained by adding an infinitesimal, $\mathcal{PT}$ symmetry breaking Zeeman energy term $\pm\eta \sigma_z$ to $h_0$ when solving Eq.~(\ref{spw}), where $\eta $ is a positive infinitesimal number. For concreteness, the  states determined by  $+\eta \sigma_z$ ($-\eta\sigma_z$) are denoted by $\bphi^+_{m,\bk}$ ($\bphi^-_{m,\bk}$) and are referred to as the spin-up (spin-down) polarized states. As a result of the $\mathcal{PT}$ symmetry these states are related to each other by ${\bm\phi}^+_{m,\bk} =\mathcal{PT}  {\bm\phi}^-_{m,\bk}$.

For weak atomic interactions and at zero temperature, the atoms naturally condense at $\bk = 0$ of the Brillioun zone ($\Gamma$ point);  the specific condensate wave function can be determined by minimizing the Gross-Pitaevskii (GP) functional with respect to the trial wave function $\bpsi(\br)$ consisting of superpositions of ${\bm\phi}_{n 0}(\br)$ from different bands. Namely, we minimize the following GP functional 
\begin{align*}
E[\bpsi]=\int d\bm r \left [ {\bm\psi}^{\dagger} h_{0} {\bm\psi } + \frac{\bar g}{2} (\bpsi^\dag \bpsi)^2 +\frac{\delta g}{2} (\bpsi^\dag\sigma_z\bpsi)^2\right ]
\end{align*}
under the normalization $ \int d\br  \bpsi^\dag (\br) \bpsi(\br) = N$, where $N$ is the total number of the atoms. The normalized condensate wave function so obtained will be denoted by ${\bm \Phi} =  (\Phi_\uparrow ,\Phi_\downarrow )^T$, for which  $\int d\br \bPhi^\dag (\br) \bPhi(\br) =1$. 

  \begin{figure}[htbp]
	\centering
	\includegraphics[width=8.5cm]{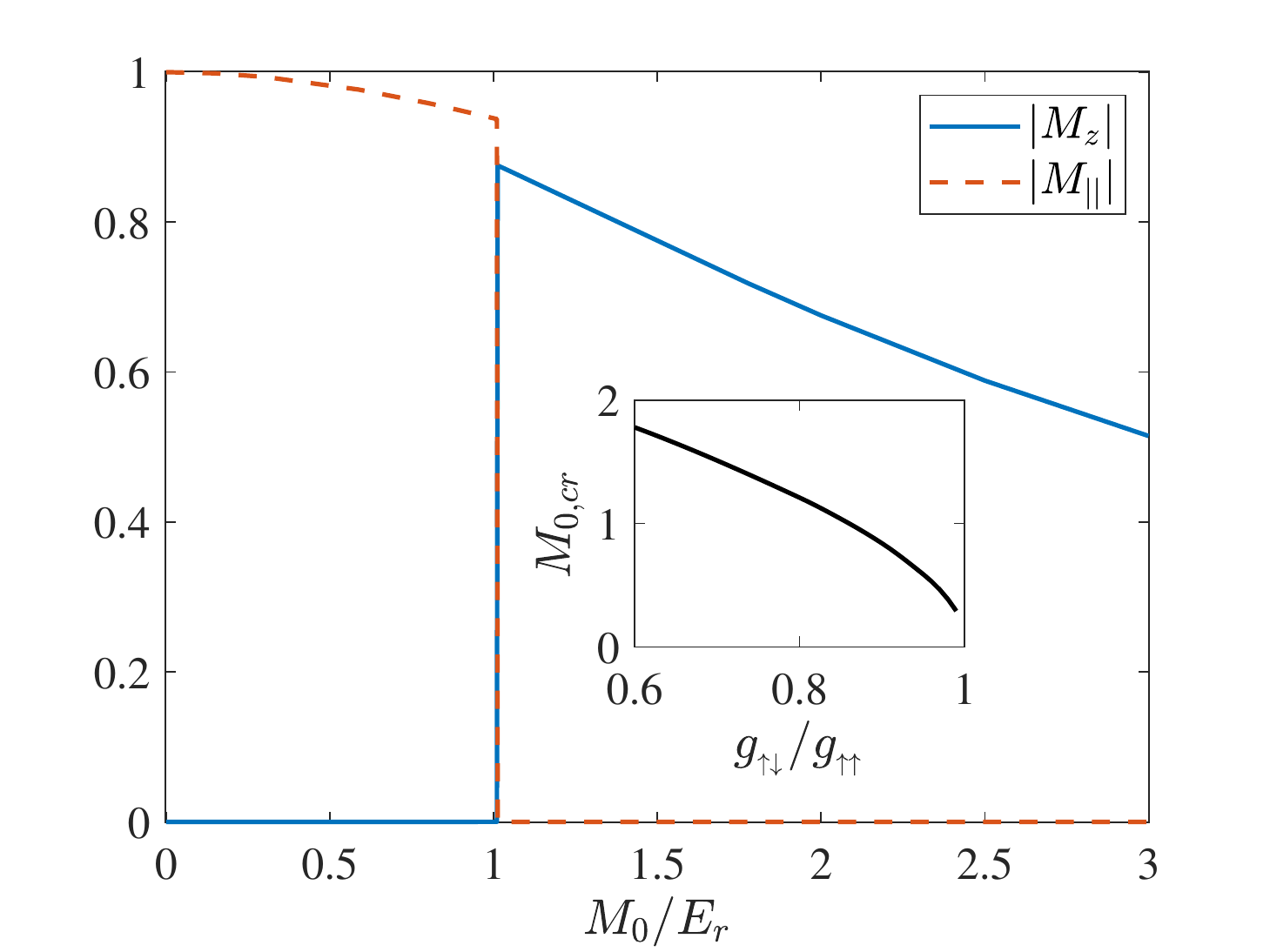}
	\caption{Magnetization as a function of the SO coupling strength $M_0$. The inset shows the critical $M_{0,cr}$ as a function of the interaction anisotropy $g_{\uparrow\downarrow}/g_{\uparrow\uparrow}$.   } 
	\label{fig3}
\end{figure}

\begin{figure*}[ht]
	\centering
	\subfigure[\ $|\Phi_\uparrow(\br)|^2$]{
	\includegraphics[width=0.2\textwidth]{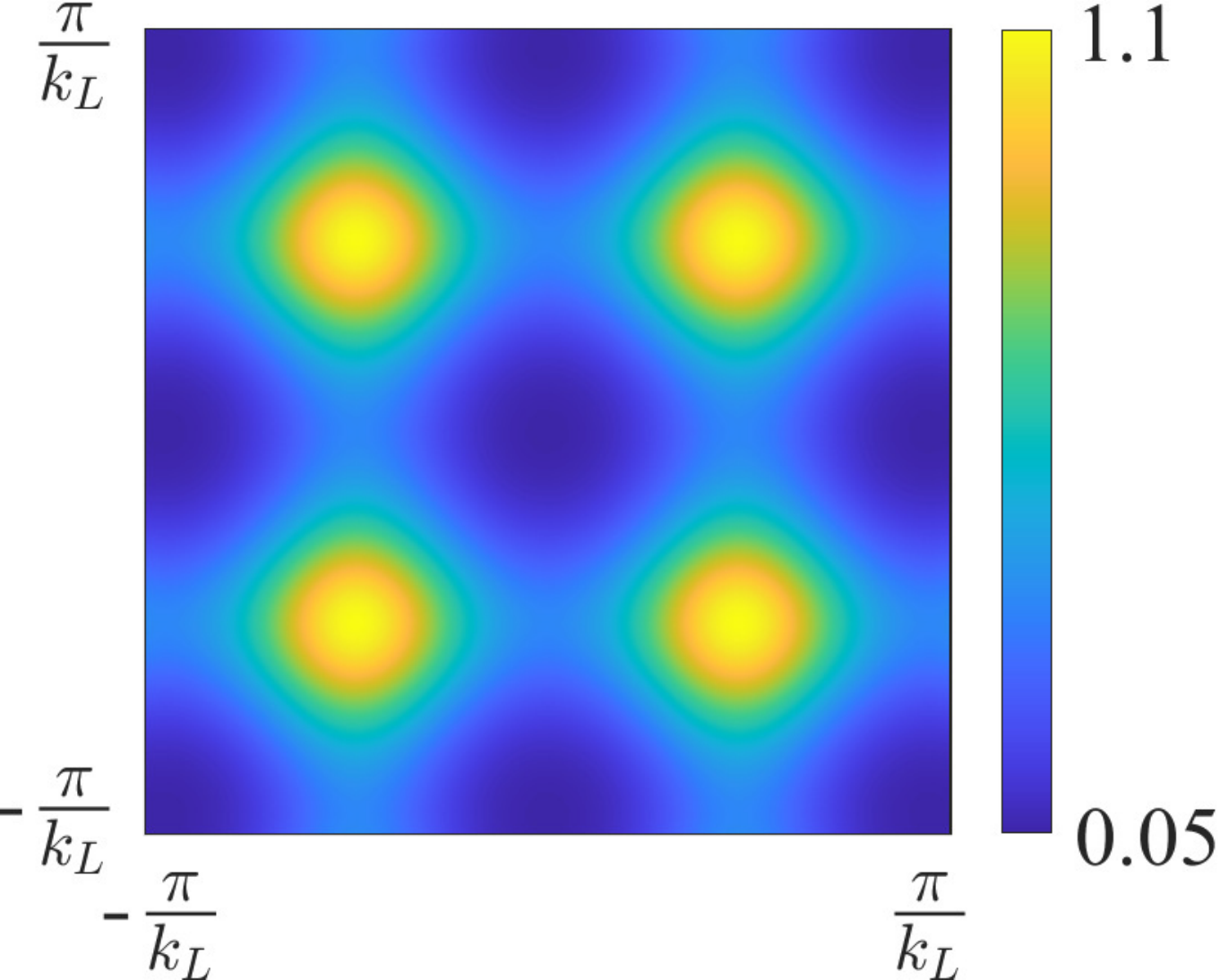}}
		\subfigure[\ $\arg \Phi_\uparrow (\br) $]{
	\includegraphics[width=0.18\textwidth]{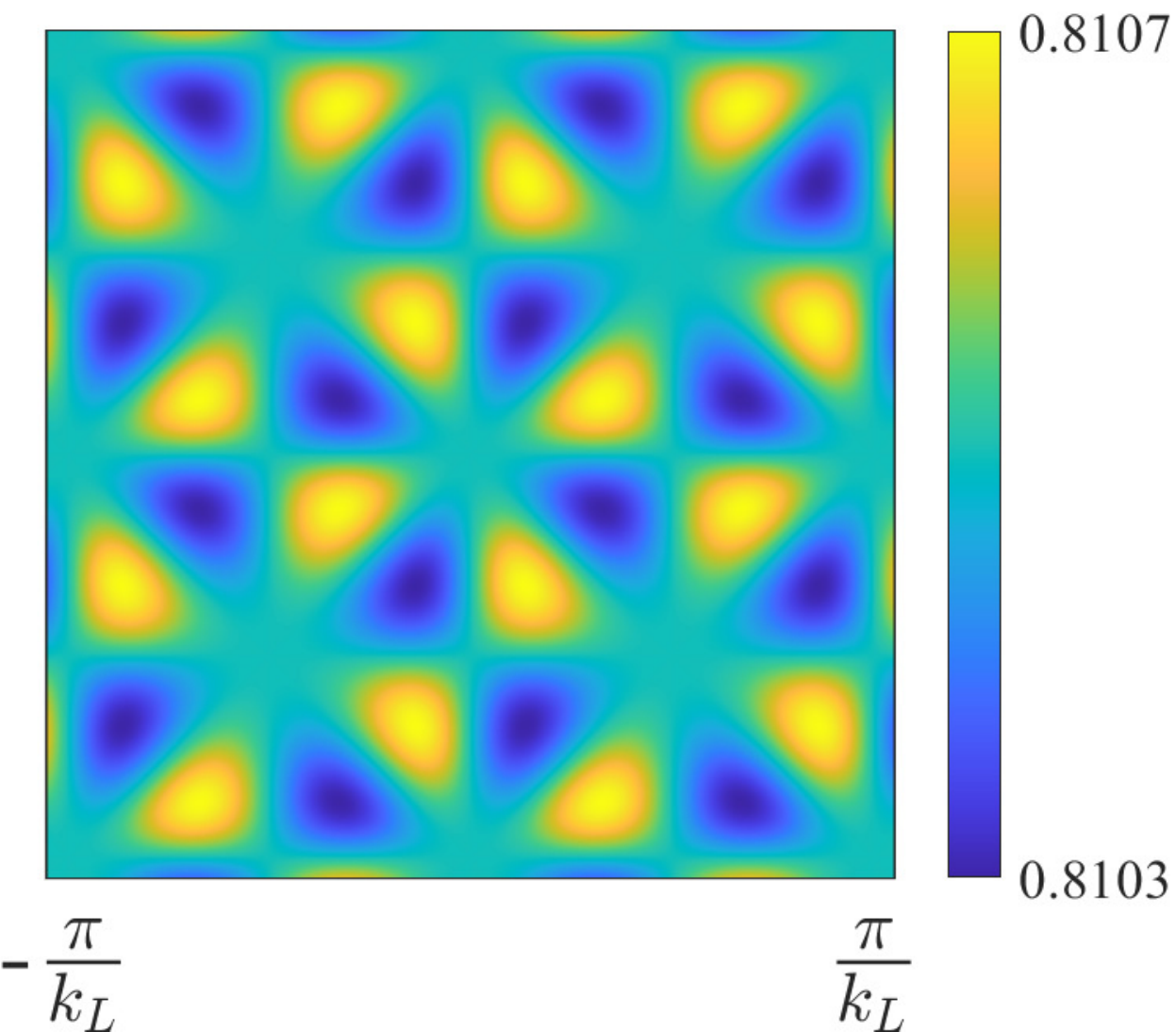}}
	\subfigure[\ $|\Phi_\downarrow(\br)|^2$]{
	\includegraphics[width=0.18\textwidth]{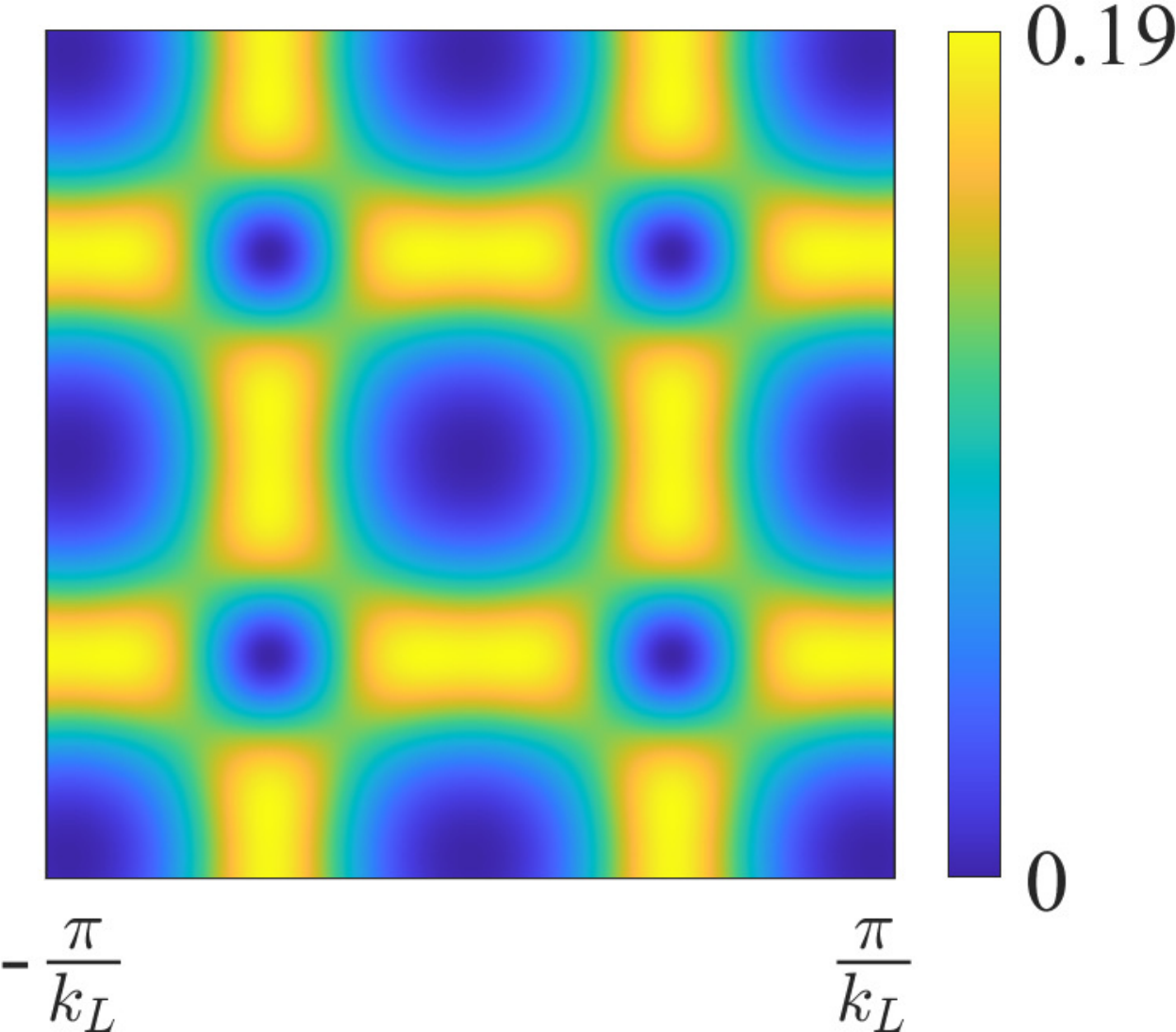}}
	\subfigure[\ $\arg \Phi_\downarrow (\br) $]{
	\includegraphics[width=0.18\textwidth]{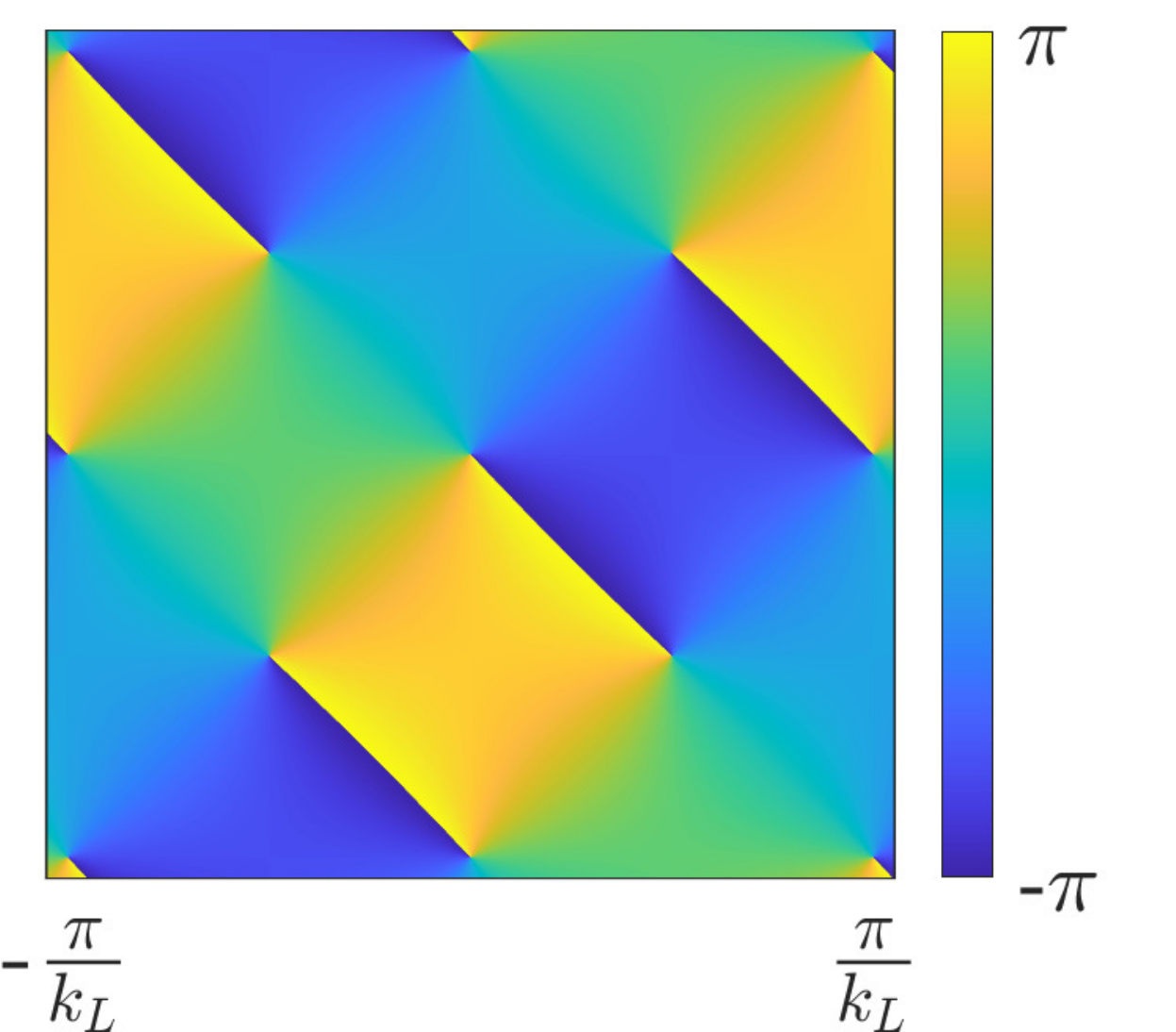}}
	\centering
	\subfigure[\ $\boldsymbol m (\br)$ ]{
	\includegraphics[width=0.18\textwidth]{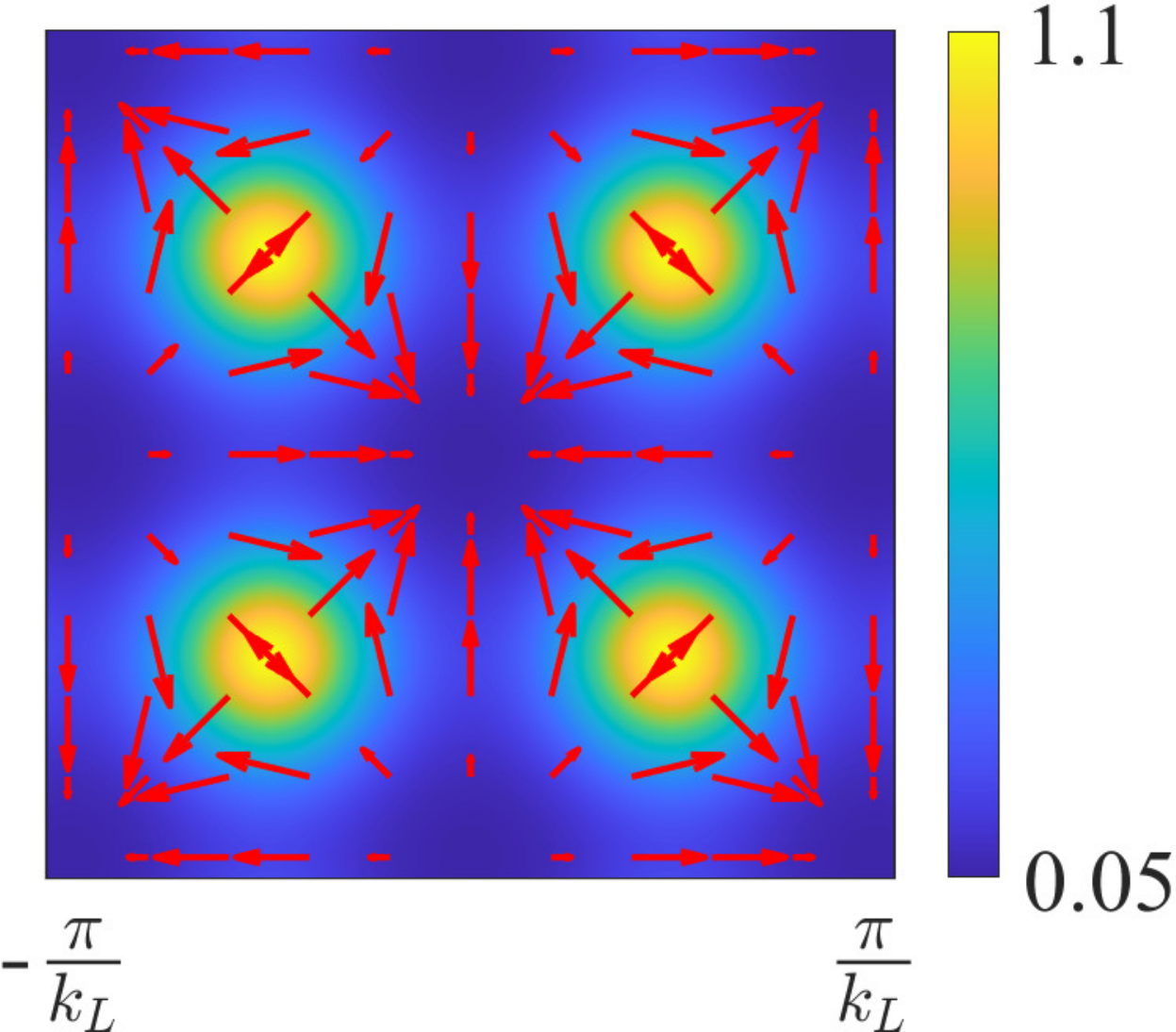}}
	\subfigure[\ $|\Phi_\uparrow(\br)|^2$]{
	\includegraphics[width=0.2\textwidth]{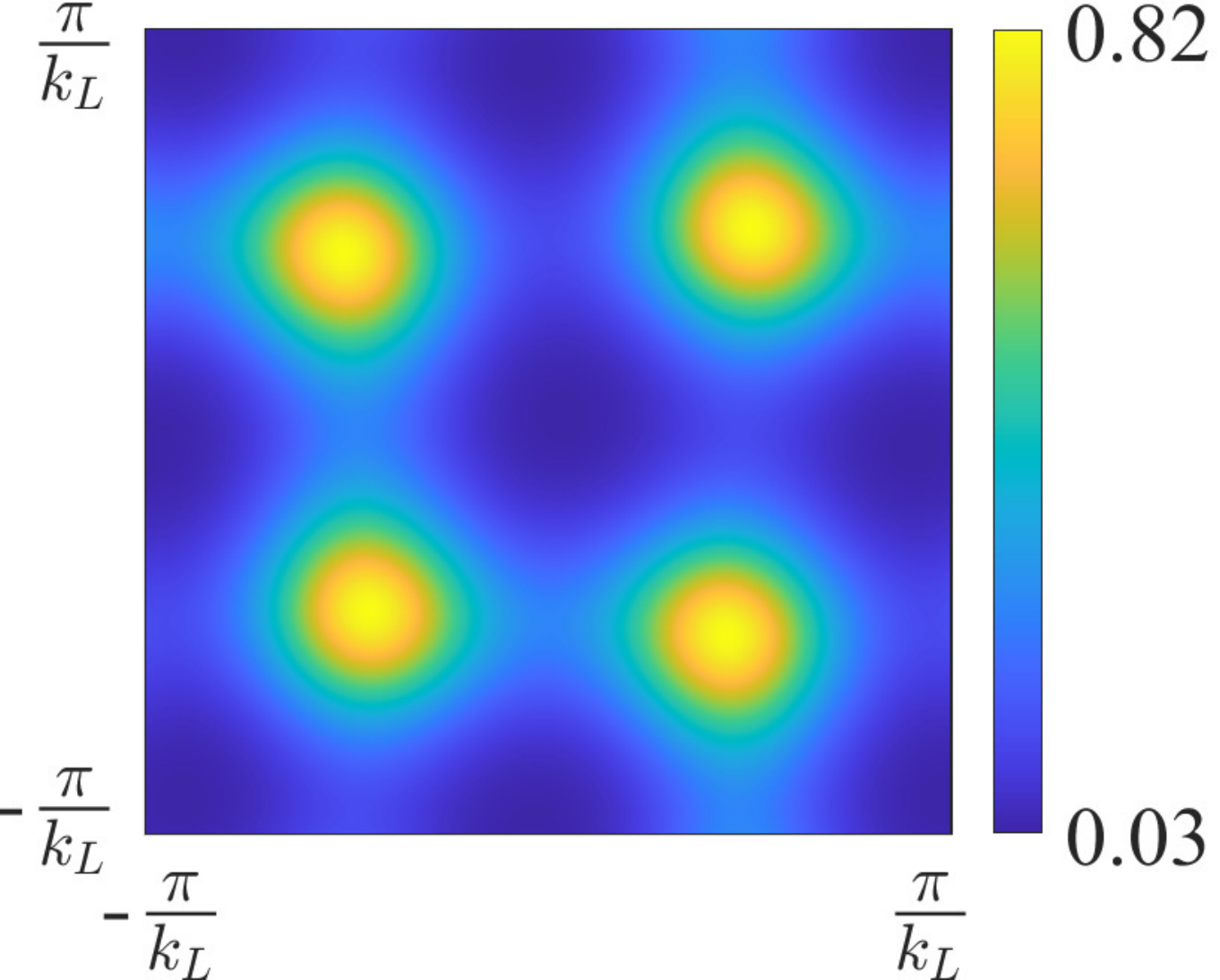}}
		\subfigure[\ $\arg \Phi_\uparrow (\br) $]{
	\includegraphics[width=0.18\textwidth]{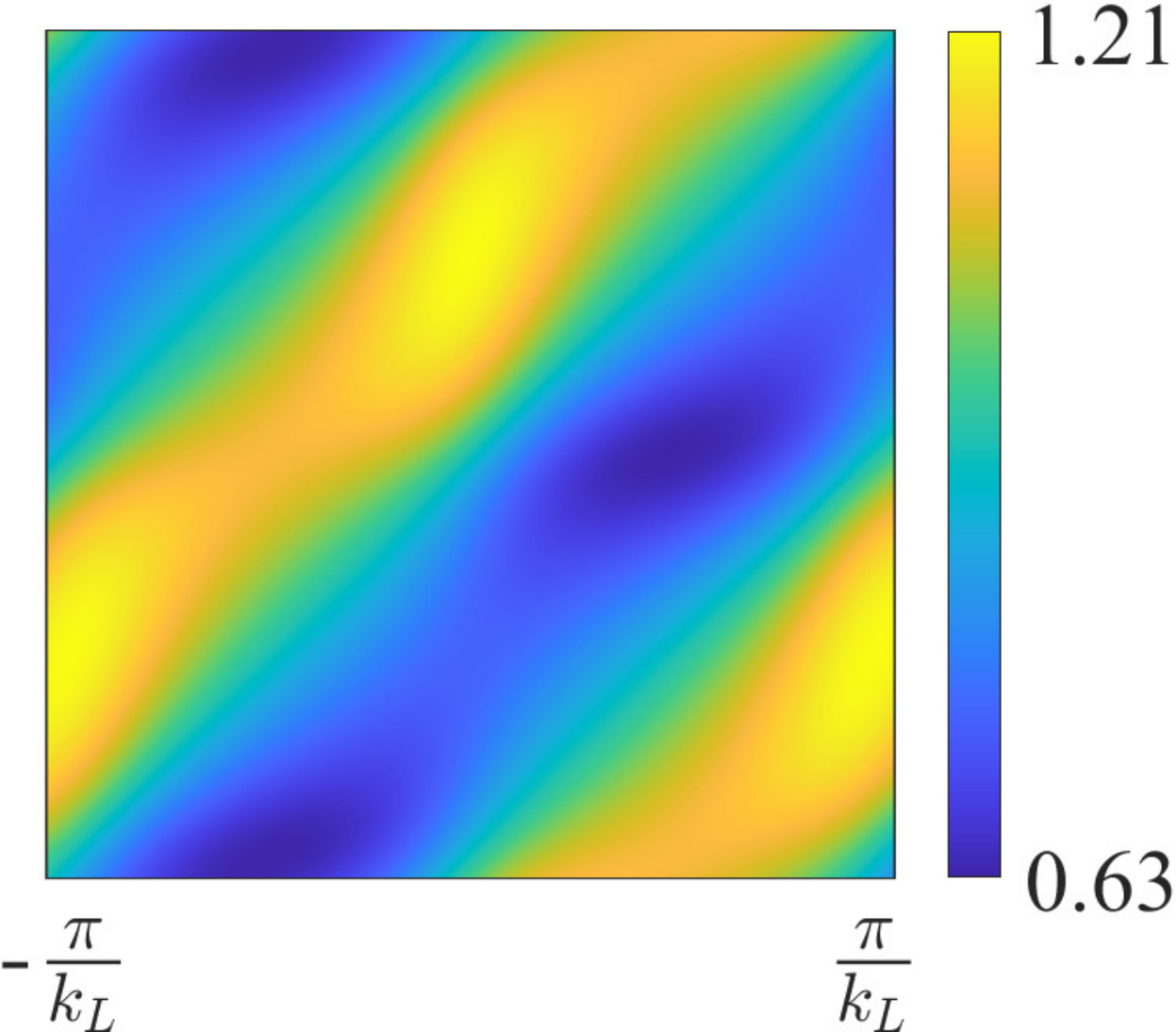}}
	\subfigure[\ $|\Phi_\downarrow(\br)|^2$]{
	\includegraphics[width=0.18\textwidth]{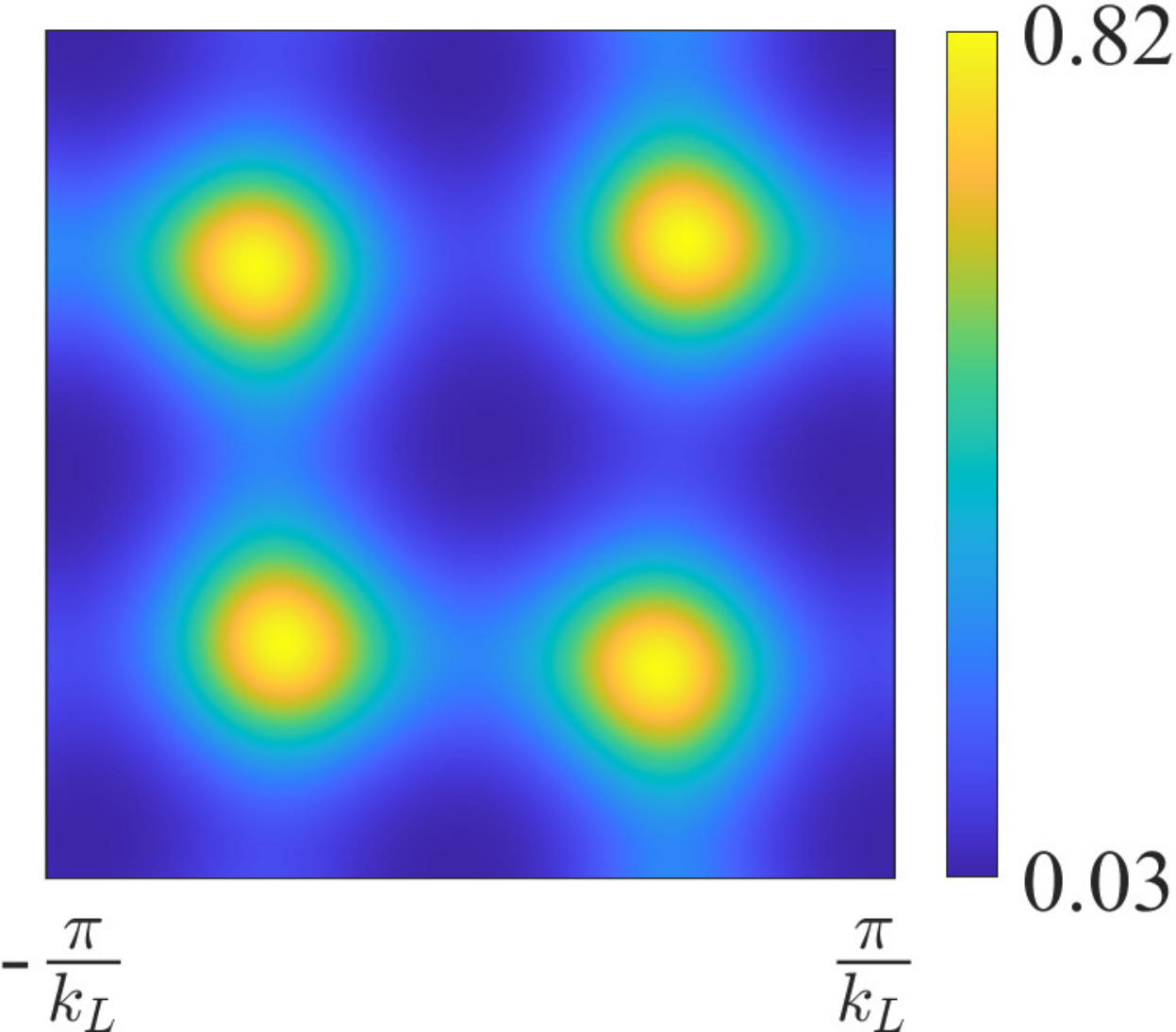}}
	\subfigure[\ $\arg \Phi_\downarrow (\br) $]{
	\includegraphics[width=0.18\textwidth]{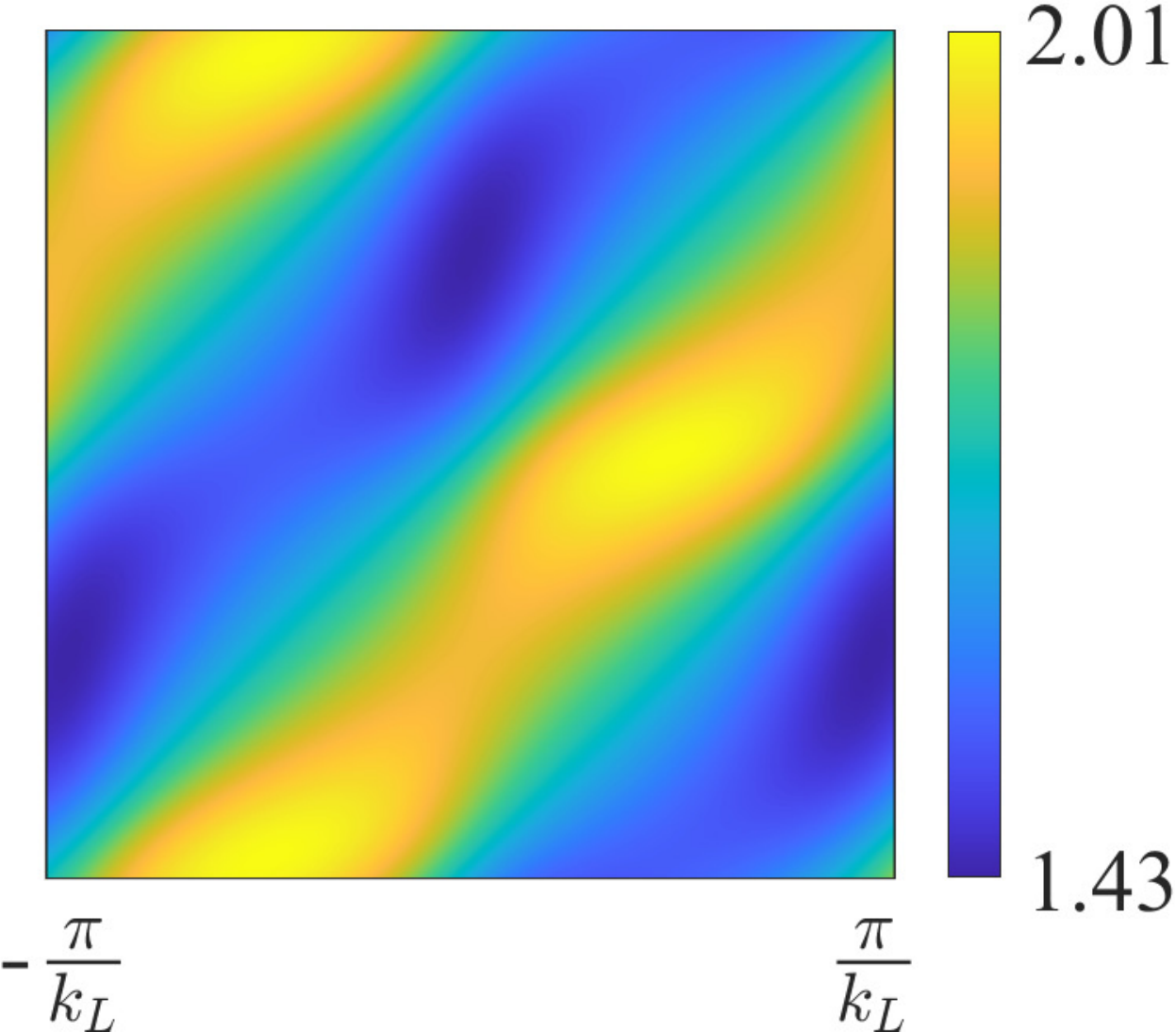}}
	\subfigure[\ $\boldsymbol m(\br)$ ]{
	\includegraphics[width=0.18\textwidth]{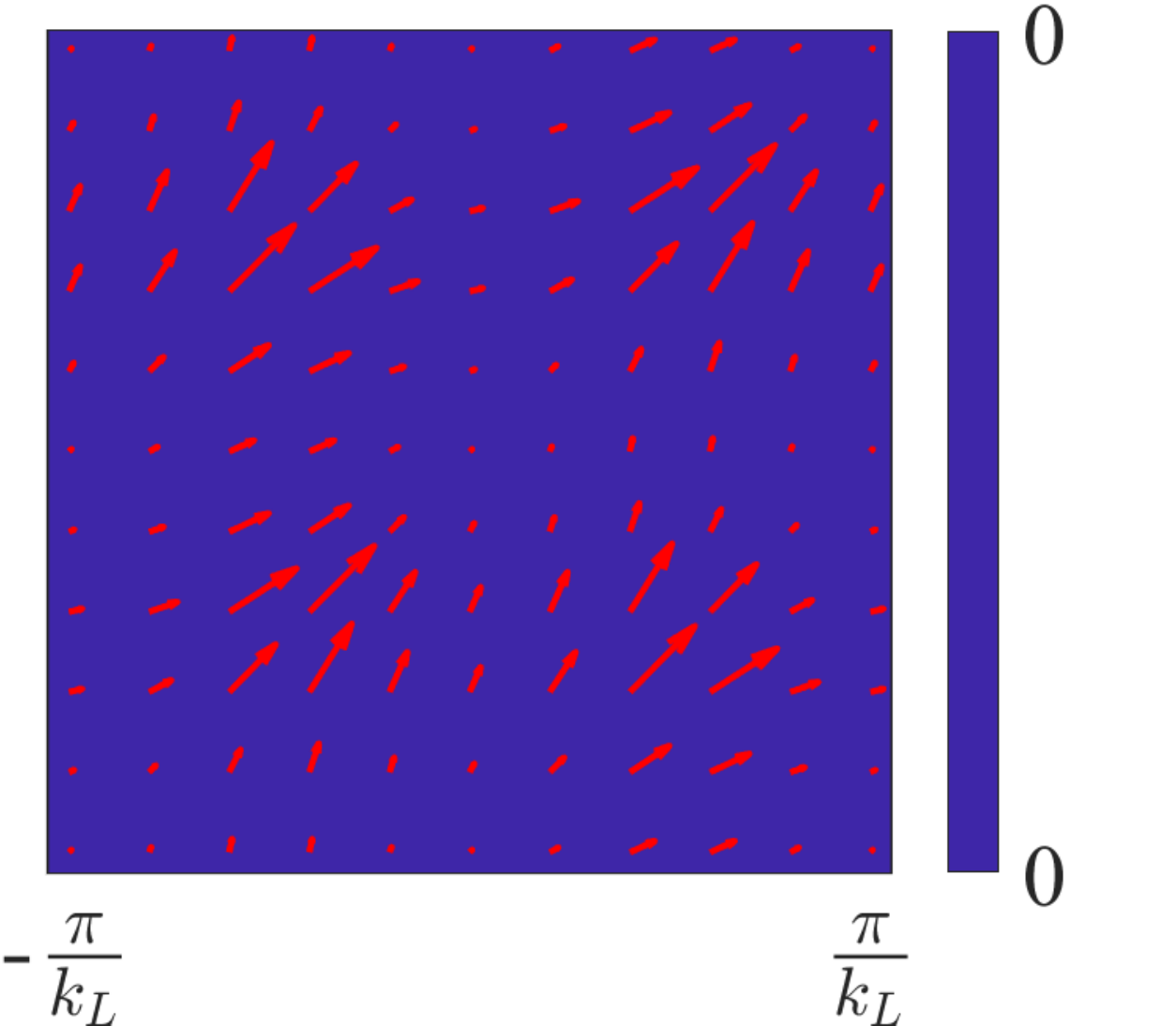}
	}
	\caption{Condensate wave function $\bPhi(\br)$ (normalized within the Wigner-Seitz cell) and magnetization density $\boldsymbol m(\br)$. Upper panel: perpendicular magnetization phase with SO coupling strength $M_0 = 2.5 E_r$; Lower panel: in-plane magnetization phase with SO coupling strength $M_0 = 0.8 E_r$. Other parameters are $V_0 = 4 E_r$, $\rho g_{\uparrow\uparrow}=0.35E_r$ and $\rho g_{\uparrow\downarrow}=0.3E_r$. These values will be used in the calculations throughout the paper unless indicated otherwise. }
	\label{fig4}
\end{figure*} 

To describe the magnetic properties we define the  magnetization per particle 
 \begin{align}
  {\boldsymbol M}  =\int d\br \, \boldsymbol m (\br) =  \int d\br \, {\bm \Phi}^\dag(\br) 
{\bm \sigma} {\bm \Phi}(\br),
 \end{align} 
where ${\bm \sigma} \equiv (\sigma_x\, ,\sigma_y,\,\sigma_z)$. For weak SO coupling strengths, the system favors a ground state which minimizes the interaction energy term associated with spin polarization, $\frac{\delta g}{2}\int d\br (\bpsi^\dag\sigma_z\bpsi)^2 $. Recalling that $\delta g = (g_{\uparrow\uparrow} - g_{\uparrow\downarrow} )/2 > 0$, we then expect a ground state with $M_z = 0$. As the SO coupling strength increases, the Raman potential energy may be gradually lowered by the spin flip induced by the SO coupling term; the amount lowered eventually compensates for the increase of the interaction energy, leading to a state with a finite magnetization along the $z$ direction. This picture is indeed confirmed by our calculation. In Fig.~\ref{fig3} we plot the magnitude of  $|M_z|$  as a function of the SO coupling strength $M_0$ for $V_0 = 4 E_r$, $\rho g_{\uparrow\uparrow} = 0.35 E_r$ and $\rho g_{\uparrow\downarrow} = 0.3 E_r$, where $\rho = N/\calA$ is the density per unit area. We see that a transition from a phase with a vanishing $M_z $ to that with a finite $M_z$  occurs at a  SO coupling strength $M_{0,cr}\approx E_r$.  The critical value of SO coupling strength depends significantly on the degree of interaction anisotropy, as shown in the inset of Fig.~\ref{fig3}.  Interestingly,  the former phase is in fact not characterized by a vanishing  total magnetization but rather by a finite in-plane magnetization$|M_{\parallel}| \equiv \sqrt{M_x^2 + M_y^2}$. In other words, our calculations show a transition from in-plane magnetization to perpendicular magnetization driven by the SO coupling strength. The properties of these two phases will be examined more closely below. 

\begin{figure}[htbp]
	\centering
	\includegraphics[width=8.5cm]{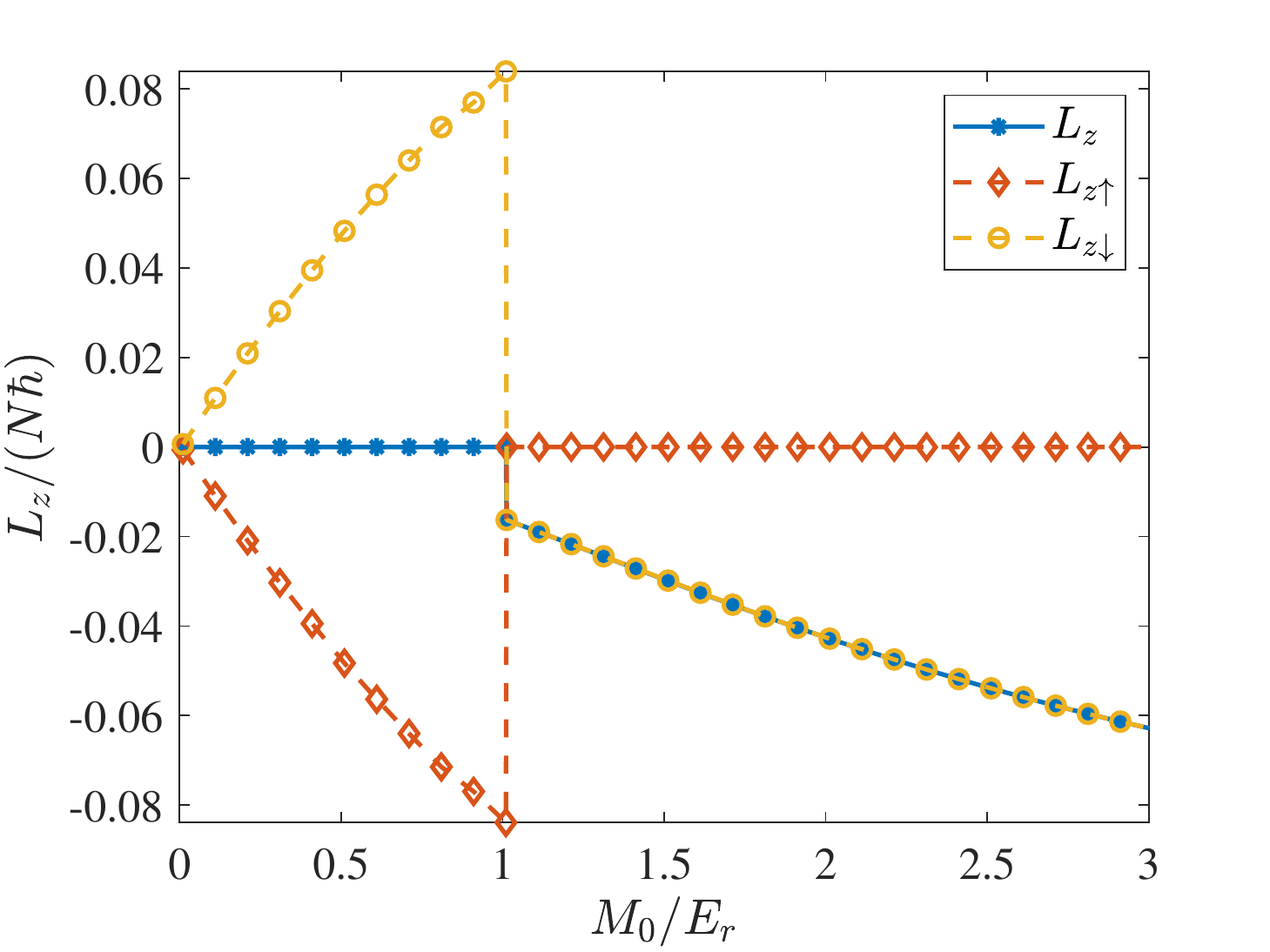}
	\caption{Angular momentum of the SOC condensate as a function of the SO coupling strength $M_0$. Here the center of the Wigner-Seitz cell is chosen as the origin in the calculation.} 
	\label{fig5}
\end{figure}

In the perpendicular magnetization phase, the ground state has a two-fold degeneracy as the direction of  the magnetization can be either along the $+z$ or the $-z$ direction. For $M_z > 0$ the condensate wave function is given by $ {\bm\Phi}=\sum_{m} \left (c^+_m {\bm\phi}^+_{m0} + c^-_m {\bm\phi}^-_{m0} \right) $ where the ${\bm\phi}^+_{m0} $ components are dominant. Clearly, the ground states with opposite $M_z$ are related to each other by the $\mathcal{PT}$ symmetry operation.  Let's take a specific value of $M_0 = 2.5 E_r$ as an example and consider the case of $M_z > 0$. The spin-up and spin-down components of this condensate wave function are shown in Fig.~\ref{fig4}(a)-(d). The most conspicuous property of the wave function is that the spin-down component contains a significant mixture of the $p$-orbitals of the square optical lattice, which is clearly reflected by the phase winding shown in Fig.~\ref{fig4}(d). In fact the spin-down component forms a vortex lattice with the vortex cores located at the optical lattice sites (see Fig.~\ref{fig4}(c)).  The spin-up component, on the other hand, consists of dominantly $s$-orbitals; its phase variation shown in Fig.~\ref{fig4}(b) indicate very small mixtures of higher orbitals. A direct implication of the vortex lattice is that the condensate carries a macroscopic angular momentum, which can in principal be calculated by
\begin{align}
L_z = N \sum_{\sigma} \int d\br  \Phi^\dag_\sigma (\br )\left (xp_y - y p_x\right ) \Phi_\sigma (\br ).
\end{align}
As is expected, the angular momentum is mostly carried by the spin-down component in this case which is shown in Fig.~\ref{fig5}. The fact that $L_z < 0 $ is consistent with the phase winding of the wave function shown in Fig.~\ref{fig4}(d).

In addition, from the condensate wave function shown in Fig.~\ref{fig4}(a)-(d), we find that the ground state breaks the $\tilde D_4$ symmetry but still retains the $\tilde C_4$ symmetry, the latter of which is formed by all the ${\tilde r_n}$ operations. It can be shown that as a consequence the magnetization density $\boldsymbol m(\br) $ must have the $C_4$ symmetry and $M_\parallel = 0$, which are confirmed by the calculation shown in Fig.~\ref{fig4}(e). Finally, the condensate wave function also preserves the nonsymmorphic symmetry $\Lambda_i$ and $\Lambda_i R$  for $R \in \tilde C_4$. Specifically we find 
\begin{equation}
\Lambda_{x} {\bm \Phi} =  - i \bPhi; \quad \Lambda_{y} {\bm \Phi} =  -i \bPhi.
\end{equation}

Now we turn to the in-plane magnetization phase, where the condensate wave function is found to be a superposition of $+\eta \sigma_z$ and $-\eta \sigma_z$ states with equal weight, namely ${\bm\Phi}=\sum_{m} \left ( c^+_m {\bm\phi}^+_{m0}+c^-_m {\bm\phi}^-_{m0}\right )$ with $|c^+_{m}|=|c^-_{m}|$. Our calculations show that there are four degenerate ground states in this phase, characterize by four possible directions of the in-plane magnetization, i.e., along $\pm \vec s_1$ and along $\pm \vec s_3$. The condensate wave function for these states can be distinguished by the relative phase of the dominant coefficients $c_1^+$ and $c_1^-$, which assume values of $(2q-1)\pi/4 $ with $q = 1,2,3, 4$. As a specific example, the condensate wave function corresponding to $q=1$ is shown in Fig.~\ref{fig4}(f)-(i).  
From these results, we find that wave function is invariant under the symmetry group $\tilde D_1 = \{ e=\tilde r_8, \tilde s_1, \tilde s_1^2 = \tilde r_4, \tilde s_1^3 = \tilde s_5\}$.  Indeed, it can be confirmed that the wave function satisfies 
\begin{align}
\tilde s_1 {\bm\Phi} = -i {\bm\Phi}.
\end{align} 
This allows us to show that the direction of the magnetization is along the $\vec s_1$ axis, as can be seen in Fig.~\ref{fig4}(j). In addition, because 
\begin{align}
\tilde{s}_{1}\hat L_{z}\tilde{s}_{1}^{-1} = -\hat L_{z},
\end{align}
where $\hat L_z =\sum_\sigma \int d\br \hat\psi^\dag_\sigma(\br) (xp_y - y p_x)\hat\psi_\sigma(\br)$, the total angular momentum of the state vanishes due to this symmetry. 
This is also confirmed by the calculations shown in Fig.~\ref{fig5} where we see that the up and down component carries opposite angular momentum. Lastly, the condensate wave function preserves the nonsymmorphic symmetries $\Lambda_i \tilde r_2$ and $\Lambda_i \tilde r_2 R$ for $R\in \tilde D_1$, as can be easily checked that 
\begin{equation}
\Lambda_x \tilde r_2 \bPhi = -\bPhi; \quad \Lambda_y \tilde r_2 \bPhi = - \bPhi.
\end{equation}

\section{Finite-frequency anomalous Hall effect} 
\label{FAHE}
Previous calculations of the ground state properties reveal that in the perpendicular magnetization phase the SOC gas contains a vortex lattice in one of the spin components and consequently carries a global angular momentum. This is reminiscent of the bosonic chiral superfluid in the boron nitride lattice where an intrinsic AHE was predicted recently~\cite{Huang2022}. Motivated by this observation, we now explore the prospects of the anomalous Hall effect in the SOC system and calculate the frequency-dependent Hall conductivity by means of the Kubo formula and the Bogoliubov theory. First, we make use of the spectral representation of the Kubo formula, which expresses the Hall conductivity in terms of the energies of the collective excitations and the transition matrix elements of the current operator. These latter quantities are then calculated using the Bogoliubov theory, aided by group symmetry analysis.  We find that the current matrix elements obey certain selection rules determined by the symmetry properties of the Bogoliubov Hamiltonian, which in turn are dependent on those of the condensate wave function. These selection rules can be used to understand the overall structure of the frequency-dependent Hall conductivity and to explain the existence of the dc AHE in the perpendicular magnetization phase and the absence of it in the in-plane magnetization phase. 
\subsection{Hall conductivity: spectral representation}
The Kubo formula for the Hall conductivity is 
\begin{align}
\sigma_{H}(\omega)\equiv  -\frac{i}{\mathcal A\omega}\chi^J_{x,y}(\omega),
\label{Kubo}
\end{align}
where $\mathcal A$ is the  system's area, $\chi^O_{x,y}(\omega)$ is the Fourier transform of the retarded correlation function of operator $ \boldsymbol { \hat O}$ 
\begin{align}
\chi^O_{x,y}(t-t') = -i\hbar^{-1} \theta(t-t') \la {[\hat O_x(t), \hat  O_y(t')]} \ra,
\end{align} and the total current operator is 
\begin{align}
\hat \bJ&=  \frac{\hbar}{ m i} \sum_\sigma \int d\br \hat\psi^{\dagger}_\sigma (\br) \nabla \hat\psi_\sigma (\br).
\label{current}
\end{align}
Quite generally, the Hall conductivity can be written in a form of spectral representation as
\begin{align}
 \sigma_H(\omega) =&-i\frac{1}{\calA\omega\hbar^2}\sum_{n\neq 0}\left[\frac{J_{x,0n}  J_{y,n0}  }{\hbar\omega - E_{n}+E_{0}+i0^+} \right.\nn \\
 &\left. -\frac{J_{y,0n}  J_{x,n0} }{\hbar\omega +E_{n}- E_{0}+i0^+}\right ],
 \label{Hall}
\end{align} 
where $E_n$ denotes the energy of the many-body eigenstate $|n\ra$ and  $J_{i,0n} = \la 0| \hat J_i |n\ra$ denotes the current matrix element between the ground state and the excited state $|n\ra$. 

As can be seen in Eq.~(\ref{Hall}), the Hall conductivity is a complex quantity; its real and imaginary parts are related to each other by the Kramer-Kronig relation
\begin{align}
{\rm Re}  \sigma_H(\omega) = \frac{1}{\pi} \mathcal{P}\int_{-\infty}^\infty d\omega' \frac{{\rm Im}  \sigma_H(\omega')}{\omega'- \omega}.
\end{align} Writing the matrix product $ J_{x,0n}  J_{y,n0} $ in terms of its real and imaginary part as 
\begin{align}
 J_{x,0n}  J_{y,n0}  =  I_n + i I'_n,
 \label{Jmp}
 \end{align} 
 where $I_n$ and $I_n'$ are now real,  we can obtain explicitly  the real and imaginary parts as
 \begin{align}
&{\rm Re}\,  \sigma_H(\omega) =\frac{1}{\calA\omega\hbar^2}\sum_{n\neq 0}  \bigg \{ \frac{2 \hbar\omega}{(\hbar\omega)^2 - (E_{n}-E_{0})^2}I'_n  \nn \\
& +\pi  \big [ \delta (\hbar \omega+ E_n - E_0) -\delta (\hbar \omega- E_n + E_0)\big  ]   I_n  \bigg \},
\label{Resigma}
\end{align}
and
\begin{align}
&{\rm Im}\, \sigma_H(\omega) =-\frac{1}{\calA\omega\hbar^2}\sum_{n\neq 0}  \bigg \{ \frac{2 (E_n - E_0) }{(\hbar\omega)^2 - (E_{n}-E_{0})^2}I_n  \nn \\
& +\pi  \big [ \delta (\hbar \omega+ E_n - E_0) +\delta (\hbar \omega- E_n + E_0)\big  ]   I'_n  \bigg \}.
\label{Imsigma}
\end{align}
Later we shall see that the symmetry of the Hamiltonian places various constraints on the current matrix elements. For the purpose of such symmetry analysis, it is useful to write the real and imaginary parts of the matrix elements product in Eq.~(\ref{Jmp}) in slightly different forms. Introducing the operators
\begin{align}
\hat J_{\pm} =\frac{1}{2} \left ( \hat J_x \pm i \hat J_y \right ),
\end{align}
we arrive at 
\begin{align}
\label{I0n}
I_n& =-i\left (  J_{+,0n}J_{+,n0} - c.c.\right  )\\
I'_n &= -\left ( |J_{+,n0}|^2 - |J_{- ,n0}|^2 \right ),
\label{Ip}
\end{align}
where $J_{\pm,nm} \equiv \la n | \hat J_\pm |m\ra$. 
We note that the dc conductivity, defined by ${\rm Re}\, \sigma_H(0)$, is given by
\begin{align}
 {\rm Re}\,  \sigma_H(0) =-\frac{2}{\calA\hbar}\sum_{n\neq 0}   \frac{I'_n }{(E_{n}-E_{0})^2}.
 \label{ReS0}
\end{align} 

\subsection{Bogoliugov theory}
For the SOC condensate, the above matrix elements will be evaluated using the Bogoliubov approximation, under which the current operator is given by 
\begin{align}
\hat \bJ \approx \frac{\hbar\sqrt{N}}{ m i} \sum_\sigma \int d\br \left [ \Phi^{*}_\sigma  \nabla \delta \hat\psi_\sigma  -  \Phi_\sigma  \nabla \delta \hat\psi^\dag_\sigma   \right ].
\label{Jbogo}
\end{align}
Here the fluctuation operator $\delta \hat\psi_\sigma  $ can be expressed as
\begin{align}
\delta \hat\psi_\sigma (\br )   = \sum_{n\bk} u_{n\bk\sigma}(\br) \hat \alpha_{n\bk\sigma} -  v^*_{n\bk\sigma}(\br) \hat \alpha^\dag_{n\bk\sigma},
\label{fobogo}
\end{align}
where $n = 0,1,2\cdots$ is now the band index (to be distinguished from that used previously to denote a general excited state), $\hat\alpha^\dag _{n\bk\sigma} $ is the creation operator for the Bogoliubov quasi-particle and $u_{n\bk\sigma}$ and $v_{n\bk\sigma}$ are the Bogoliubov amplitudes. These amplitudes are determined by the Bogoliubov-de Gennes (BdG) equation
\begin{align}
\tau_z{\mathcal {H}}_B(\br) {\bm {V}}_{n\bk}(\br) = \mathcal{E}_{n\bk} {\bm {V}}_{n\bk}(\br),
\label{BdGeq}
\end{align}
where $\tau_z = \sigma_z\otimes I$, $\mathcal{E}_{n\bk}$ is the band dispersion of the quasi-particle,  ${\bm {V}}_{n\bk} \equiv \left (\bu^T_{n\bk}, -\bv^T_{n\bk} \right )^T$  is the corresponding amplitude with the normalization $\int d\br V_{m\bk}^\dag(\br)\tau_z V_{n\bk}(\br) = \tau_{z,mn}$ and $\mathcal{H}_{B}(\br)$ is the Bogoliubov Hamiltonian
\begin{equation}
		\mathcal{H}_{B}(\br)=\begin{pmatrix} \mathcal{M}+h_0 -\mu &\mathcal{N} \\ \mathcal{N}^{*} & \mathcal{M}^{*}+h_0^{*}-\mu\end{pmatrix}. \label{eq:bdgre}
\end{equation}
Here  $\mu$ is the chemical potential determined earlier along with the condensate wave function ${\bm {\Phi}}$, 
\begin{equation*}
	\mathcal{M}=\rho \calA\begin{pmatrix} 2g_{\uparrow\uparrow}|\Phi_{\uparrow}|^{2}+g_{\uparrow\downarrow}|\Phi_{\downarrow}|^{2} & g_{\uparrow\downarrow}\Phi_{\downarrow}^{*}\Phi_{\uparrow} \\
	g_{\uparrow\downarrow}\Phi_{\uparrow}^{*}\Phi_{\downarrow} & 2g_{\downarrow\downarrow}|\Phi_{\downarrow}|^{2}+g_{\uparrow\downarrow}|\Phi_{\uparrow}|^{2}
	\end{pmatrix} \label{eq:BdgM}
\end{equation*}
and
\begin{equation*}
	\mathcal{N}=\rho\calA\begin{pmatrix} g_{\uparrow\uparrow}\Phi_{\uparrow}^{2} &g_{\uparrow\downarrow}\Phi_{\downarrow}\Phi_{\uparrow} \\ g_{\uparrow\downarrow}\Phi_{\uparrow}\Phi_{\downarrow} & g_{\downarrow\downarrow}\Phi_{\downarrow}^2 \end{pmatrix}. \label{eq:BdgN}
\end{equation*}
Examples of the Bogoliubov quasi-particle spectrum obtained by solving Eq.~(\ref{BdGeq}) in the perpendicular and in-plane magnetization phases are shown in Fig.~\ref{fig6} and Fig.~\ref{fig8} respectively. We note that there is a gapless solution to the BdG equation in both phases as a result of the well-known Goldstone theorem. The corresponding eigenvector, for which $n=0$ and $\bk=0$, reproduces the GP wave function and takes the form of ${\bm {V}}_{00} = \left (\bPhi^T, -\bPhi^\dag \right )^T$. 

Since the total current conserves the crystal momentum, the relevant current matrix elements within the Bogoliubov theory are those between the ground state and the excited states with one quasi-particle of zero crystal momentum. Using Eqs.~(\ref{Jbogo})-(\ref{fobogo}) we find that the matrix elements $J_{\pm,n0} $ are given by 
\begin{align}
J_{\pm,n0}  & = \frac{\hbar\sqrt{N}}{mi}\sum_\sigma\int d\br \left (u^*_{n0\sigma}   \pa_\pm \Phi_\sigma \  + v^*_{n0\sigma}  \pa_\pm \Phi^*_\sigma \right )\nn \\
& \equiv \sqrt{N}\la \bm V_{n0}| \calJ_\pm |\bm V_{00} \ra,
\label{Jmatr}
\end{align}
where 
\begin{align}
\calJ_\pm \equiv (\hbar/m i ) {\rm diag}\left (\pa_\pm, \pa_\pm, \pa_\pm,\pa_\pm  \right ),
\end{align}
with $\pa_\pm \equiv (\pa_x \pm i \pa_y)/2$. Although it is straightforward to calculate Eq.~(\ref{Jmatr}) using the solutions from the Bogoliubov equation, such a task can be greatly simplified if we make use of the symmetry properties of the Bogoliubov Hamiltonian and its eigenstates. First, it can be shown by such symmetry analysis that the matrix elements product $J_{x,0n}J_{y,n0}$ in Eq.~(\ref{Jmp})  is purely imaginary in the perpendicular magnetization phase and purely real in the in-plane magnetization phase. This property alone indicates that the dc AHE is absent in the latter phase. Furthermore, these matrix elements obey selection rules governed by the symmetry group of the Bogoliubov Hamiltonian; these selection rules determine the overall structure of the complex Hall conductivity. 

In order to demonstrate the above properties, we first discuss the symmetry of the Bogoliubov Hamiltonian, which depends crucially on that of the condensate wave function $\bPhi$. We found earlier that the even though $\bPhi$ breaks the $\tilde D_4$ symmetry, it nevertheless preserves a symmetry subgroup of $\tilde D_4$. Suppose that $\bPhi$ is invariant under a  symmetry operation $R\in \tilde D_4$, i.e., 
\begin{align}
R \bm \Phi = e^{i\theta_R} \bm \Phi.
\label{RPhi}
\end{align}
Then it is straightforward to show that
\begin{align}
f_1(R) = \begin{pmatrix} e^{-i\theta_R} R&  \\  &\calK e^{-i\theta_R}   R \calK  \end{pmatrix}
\label{RB}
\end{align}
 is a symmetry operator of $\calH_B$, namely 
\begin{align}
f_1 (R)\calH_B f_1^{-1} (R)= \calH_B.
\label{HBinv}
\end{align}
In addition,  the condensate wave function may also retain certain nonsymmorphic symmetry $\Lambda_i R  $. In this case, we have 
\begin{align}
(\Lambda_i R) \bPhi = e^{i\beta_{i,R}}\bPhi,
\end{align}
which allows us to similarly define a corresponding symmetry operation for $\calH_B$ as
\begin{align}
f_2(\Lambda_i R ) = \begin{pmatrix} e^{-i\beta_{i,R}} \Lambda_iR&  \\  &\calK e^{-i\beta_{i,R}}   \Lambda_i R \calK  \end{pmatrix}.
\label{f2}
\end{align}
In the following, we will discuss in detail the implications of  these symmetry operators for $\calH_B$. To do so, we need to treat the two magnetic phases separately because their corresponding condensate wave functions possess different symmetry subgroups of $\tilde D_4$ and also different nonsymmorphic symmetries. 

\subsection{Perpendicular magnetization phase }
As previously found, the condensate wave function $\bPhi$ in the perpendicular magnetization phase preserves the $\tilde C_4$ symmetry.  In Tab.~\ref{TabC4}  we display the character table of relevant 1D irreducible representations of the $\tilde C_4$ group.  We again take the case of $M_z >0$ as an example, where our calculation shows that $\bPhi$ is the basis function of the 1D irreducible representation $A_5$ of the $\tilde C_4$ group. Thus from Eq.~(\ref{RPhi}) we have
\begin{align}
e^{i\theta_R} = \chi^{(A_5)}(R), 
\end{align}
where $\chi^{(\Gamma)}(R)$ denotes the character of $R$ in the 1D  representation $\Gamma$ of $\tilde C_4$. With this phase factor so determined, it is straightforward to  show from Eq.~(\ref{RB}) that  $f_1(R)\cdot f_1(R') = f_1(R\cdot R')$ for arbitrary $ R,R' \in \tilde C_4$.  This means that  all the $f_1(R)$ operations for $ R \in \tilde C_4$ form a symmetry group of $\calH_B$ isomorphic to $\tilde C_4$, which we denote as $\tilde \calC_4 \equiv \{ f_1(R): \forall R \in \tilde C_4 \}$. Since the solutions of the BdG equation are non-degenerate, the existence of this symmetry group implies that $\bV_{n0}$ is the basis function of some 1D irreducible representation of $\tilde C_4$,  denoted by $\Gamma_n$. Each of $\Gamma_n$ can be found by calculating $f_1(R)  \bV_{n0}$ and comparing the result to the character table in Tab.~\ref{TabC4}. For example, it is easy to show that $\Gamma_0$ is the trivial representation $ A_1$. 
In addition, by direct calculations of $f_1(R)\calJ_\pm f_1^{-1}(R)$ we find that the operators $\calJ_\pm$ are also basis functions of certain 1D irreducible representations of $\tilde C_4$, denoted by $\Gamma_\pm$. Comparing these calculations to the character table of the 1D representations of $\tilde C_4 $ in Tab.~\ref{TabC4}, we find that $\Gamma_+ = A_2$ and $\Gamma_- = A_3$.

\begin{table}[htbp]
	\centering
\[
\begin{array}{c|cccccccc}
\tilde{C}_{4} & e & \tilde{r}_{1} & \tilde{r}_{2} & \tilde{r}_{3} & \tilde{r}_{4} & \tilde{r}_{5} & \tilde{r}_{6} & \tilde{r}_{7}\\
\hline 
A_{1} & 1 & 1 & 1 & 1 & 1 & 1 & 1 & 1\\
A_{2} & 1 & i & -1 & -i & 1 & i & -1 & i\\
A_{3} & 1 & -i & -1 & i & 1 & -i & -1 & i\\
A_{4} & 1 & -1 & 1 & -1 & 1 & -1 & 1 & -1 \\
A_5 & 1& e^{-i\pi/4}& e^{-i\pi/2} &e^{-i3\pi/4} &e^{-i\pi} &e^{i3\pi/4} & e^{i\pi/2} &e^{i\pi/4}  
\end{array}
\]
	\caption{Character table of $\tilde C_4$.}
	\label{TabC4}
\end{table}

\begin{figure}[htbp]
	\centering
	\includegraphics[width=8cm]{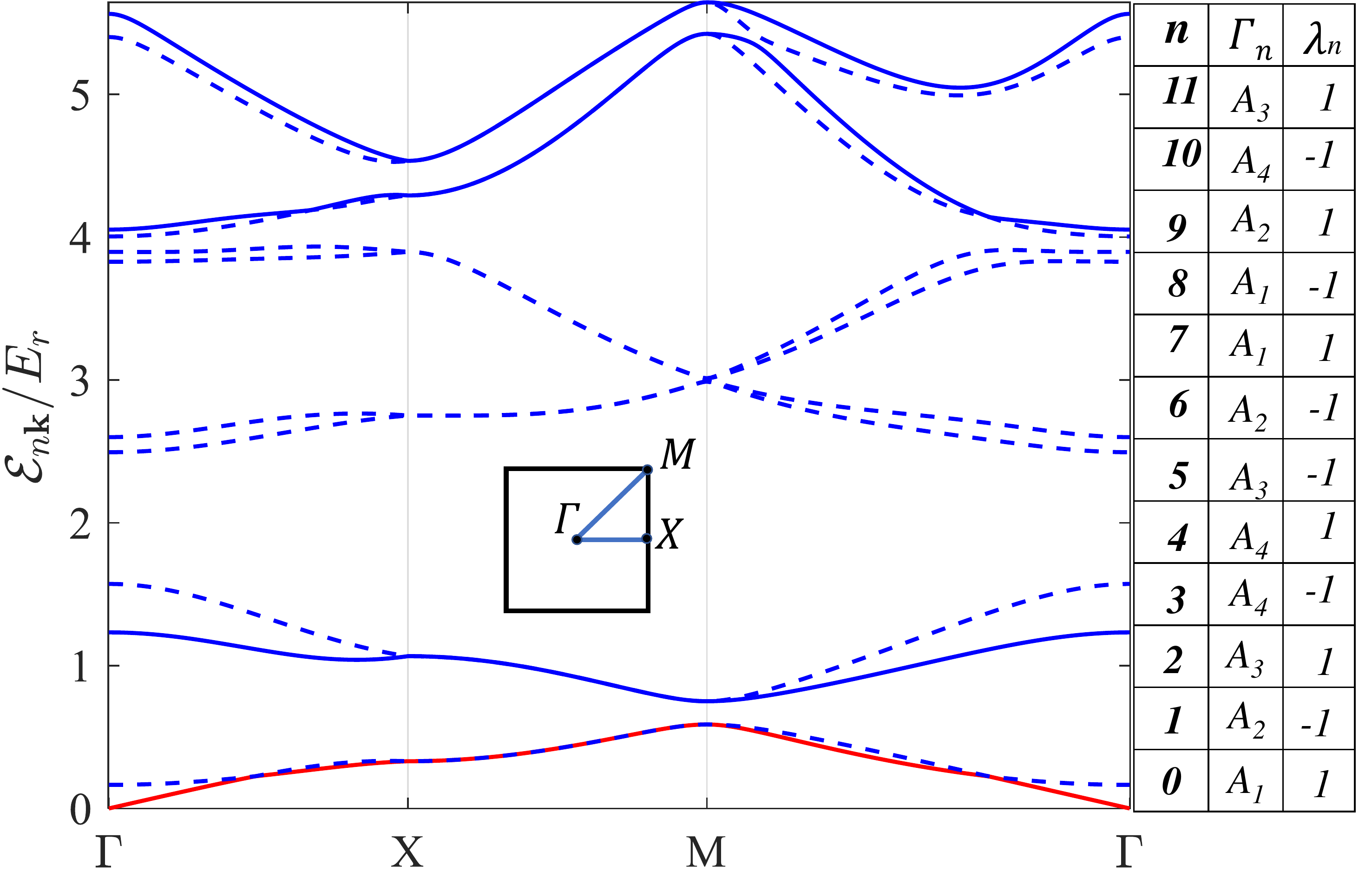}
	\caption{Bogliuobov bands in the perpendicular magnetization phase. On the right side, the $\Gamma_n$ and the $\lambda_n$ value corresponding to the $\bk = 0$ state of the $n$-th band are given. According to the selection rule  explained in the text, the transition is forbidden from the ground state to those bands indicated by dashed lines.} 
	\label{fig6}
\end{figure}
 \begin{figure}[htbp]
	\centering
	\includegraphics[width=8.5cm]{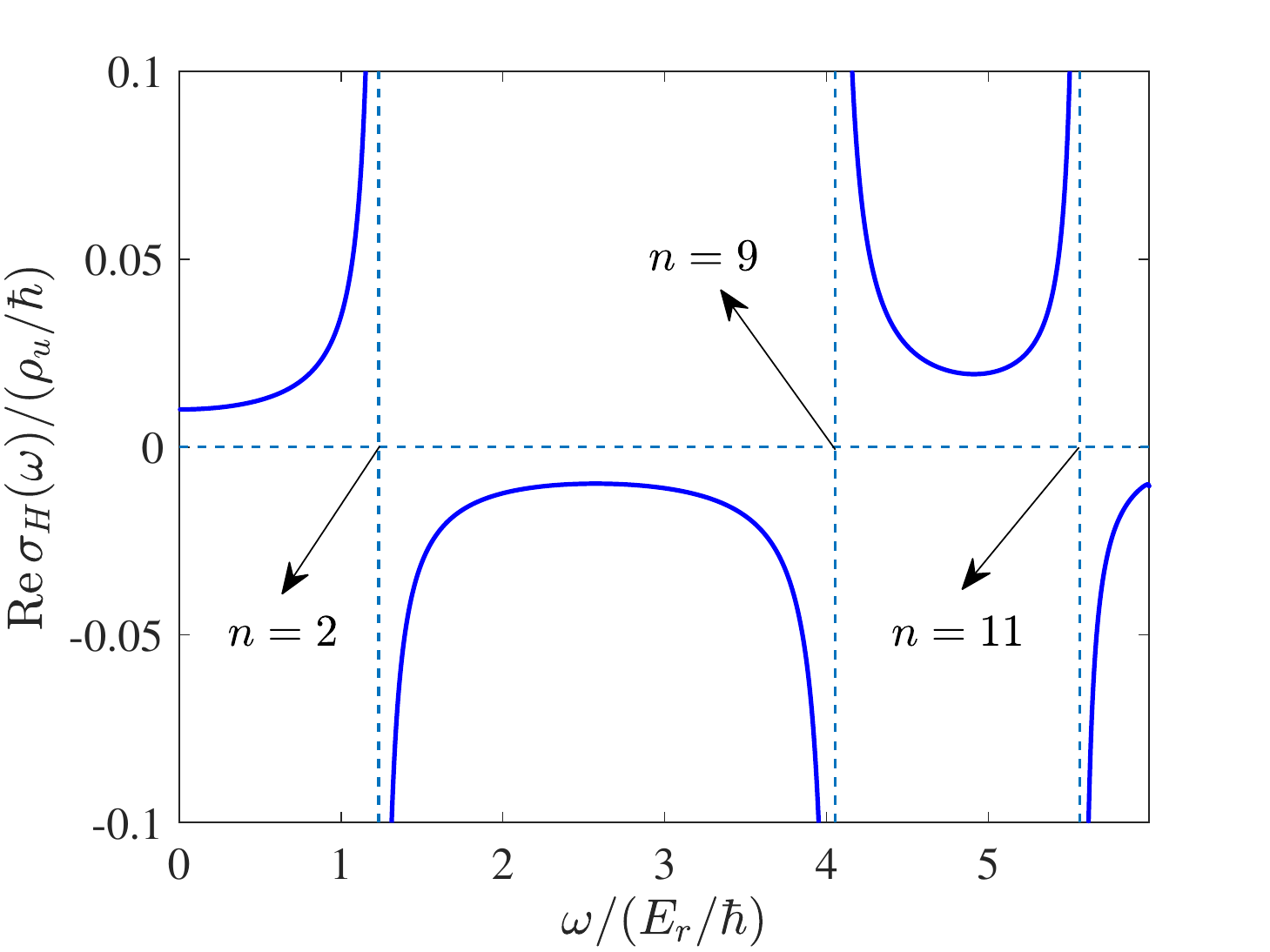}
	\caption{Real part of the frequency-dependent Hall conductivity for the system in the perpendicular magnetization phase. Here $\rho_u$ is the number of atoms per unit cell.} 
	\label{fig7}
\end{figure}

Now, we are in a position to discuss the selection rule of the matrix element $\la \bV_{n0} |\calJ_+  | \bV_{00}\ra $ required by the $
\tilde \calC_4$ symmetry group. The matrix element should transform as a constant when any symmetry group operation is applied to the states and the operator $\calJ_+$ simultaneously; otherwise it must be zero. In the language of group theory~\cite{Dresselhaus2007group}, this is saying that such a matrix element must vanish if 
\begin{align}
\Gamma_n \neq \Gamma_0\otimes \Gamma_+.
\label{J+crit}
\end{align}
Similarly we have $\la \bV_{n0} |\calJ_-  | \bV_{00}\ra  =0 $ if
\begin{align}
\Gamma_n \neq \Gamma_0\otimes \Gamma_-.
\label{J-crit}
\end{align}
The first consequence of this selection rule is that the general expressions of the Hall conductivity in Eqs.~(\ref{Resigma})-(\ref{Imsigma}) can be immediately simplified. Applying the rule to the matrix elements product $\la \bV_{00} |\calJ_+  | \bV_{n0}\ra \la \bV_{n0}| \calJ_+  |\bV_{00}\ra =  (\la \bV_{n0}| \calJ_-  |\bV_{00}\ra)^*   \la \bV_{n0}| \calJ_+  |\bV_{00}\ra  $, we see that it must vanish due to the fact that  
$
\Gamma_+ \neq  \Gamma_- 
$.
In view of Eq.~(\ref{Jmatr}) and Eq.~(\ref{I0n}), this immediately leads to $I_n = 0$. Consequently within the Bogoliubov theory the Hall conductivity takes the  form of
\begin{align}
{\rm Re} \, \sigma_H(\omega) =\frac{2}{\calA\hbar}\sum_{n\neq 0}   \frac{I'_n }{(\hbar\omega)^2 - (\calE_{n0}-\calE_{00})^2} 
\label{Resperp}
\end{align}
and
\begin{align}
&{\rm Im} \, \sigma_H(\omega) =-\frac{\pi}{\calA\omega\hbar^2} \times\nn \\
&\sum_{n\neq 0}   \big [ \delta (\hbar \omega+ \calE_{n0} - \calE_{00}) +\delta (\hbar \omega- \calE_{n0} + \calE_{00})\big  ]   I'_n,
\end{align}
where the Bogoliubov quasi-particle energy $\calE_{n0}$ is determined from the BdG equation and 
\begin{align}
I'_n = -N\left (  \left | \la \bV_{n0} |\calJ_+  | \bV_{00}\ra \right |^2 -\left | \la \bV_{n0} |\calJ_-  | \bV_{00}\ra \right |^2\right ). 
\end{align}
We note that the real part of the complex Hall conductivity is reactive while the imaginary part is absorptive, the same as those found in condensed matter systems. 
The quantity $I'_n$ is not finite for all $n$ due to the selection rule; it vanishes for those states whose corresponding 1D $\tilde \calC_4$ group representation satisfies both Eqs.~(\ref{J+crit}) and (\ref{J-crit}).  In other words, the selection rule of the $\tilde \calC_4$ symmetry  allows only transitions to states whose 1D $\tilde \calC$ group representation is either $A_2$ or $A_3$. In addition, the nonsymmorphic symmetry  imposes further restrictions among these states. Because the condensate wave function satisfies the nonsymmorphic symmetry $\Lambda_x$, a corresponding symmetry operation, $f_2(\Lambda_x)$  defined by Eq.~(\ref{f2}), exists for the Bogoliubov Hamiltonian. Under this operation  $\calJ_\pm$ and the Bogoliubov amplitude $\bV_{n0}$ transform as
\begin{align}
f_2(\Lambda_x) \calJ_{\pm} f^{-1}_2(\Lambda_x) & = \calJ_{\pm} \nn \\
f_2(\Lambda_x) \bV_{n 0}  &= \lambda_n  \bV_{n0}. 
\end{align}
This means that transitions to those states for which $\lambda_n \neq \lambda_0$ are forbidden by the nonsymmorphic symmetry.  
In Fig.~\ref{fig6} we show a table of calculated $\Gamma_n$ and $\lambda_n$ for the first $12$ bands. As can be checked from this table, the combination of the two sets of selection rules limits the relevant excited bands to $n=2,9, 11$. 

In Fig.~\ref{fig7} we have shown an example of  ${\rm Re}\, \sigma_H(\omega)$, where we find resonances  located precisely at the $\Gamma$ point excitation energies of these bands. The imaginary part  $\rm Im\, \sigma_H(\omega)$ is given by the summation of a series of delta functions centered at these excitation energies, with weights determined by the corresponding $I'_n$. Notably, our calculations show a finite dc Hall conductivity ${\rm Re}\,\sigma_H(0)$ for the perpendicular magnetization phase, which is in sharp contrast with the in-plane magnetization phase to be discussed below. We will return to the issue of dc Hall conductivity later in Sec.~\ref{DCHALL}.

\subsection{In-plane magnetization phase}
In comparison to the perpendicular magnetization phase, the condensate wave function in the in-plane magnetization phase preserves the $\tilde D_1$ symmetry. To be specific, we take the condensate wave function  considered in Sec.~\ref{PT} as an example.  Our calculation shows that this condensate wave function is the basis function of the 1D irreducible representation $B_4$ of the $\tilde D_1$ group (see the character table of the representations  of the $\tilde D_1$ in Tab.~\ref{TabD1} ). Letting 
\begin{align}
e^{i\theta_R} = \chi^{(B4)} (R),
\end{align}
we can again form a symmetry group of $\calH_B$, $\tilde \calD_1 \equiv \{f_1(R):\forall R\in\tilde D_1 \}$, which is isomorphic to $\tilde D_1$. We now show that $I'_n = 0$  as a result of this symmetry.  It can be easily checked that for $\tilde s_1 \in \tilde D_1$
\begin{align}
f_1(\tilde s_1) \calJ_+ f_1^{-1}(\tilde s_1) = i\calJ_-.
\label{EJ}
\end{align}
Indeed, it turns out that $\calJ_+$ and $\calJ_-$ are the basis functions of a 2D representation of $\tilde \calD_1$ group, denoted here by $\Gamma_\calJ$. The above transformation, together with the fact that $\bV_{n0}$ is the basis function of a 1D representation of $\tilde \calD_1$ group, leads to 
\begin{align}
 |\la \bV_{n0} |\calJ_+  | \bV_{00}\ra| =  |\la \bV_{n0} |\calJ_-  | \bV_{00}\ra|. 
 \end{align} 
In view of Eq.~(\ref{Jmatr}) and Eq.~(\ref{I0n}), we then arrive at $I'_{n} = 0$ and 
 \begin{align}
&{\rm Re} \, \sigma_H(\omega) =\frac{\pi }{\calA\omega\hbar^2}\times \nn \\ \sum_{n\neq 0} 
&  \big [ \delta (\hbar \omega+ \calE_{n0} - \calE_{00}) -\delta (\hbar \omega- \calE_{n0} + \calE_{00})\big  ]   I_n 
\label{ResigmaIP}
\end{align}
and
\begin{align}
&{\rm Im} \, \sigma_H(\omega) =-\frac{1}{\calA\omega\hbar^2}\sum_{n\neq 0}  \frac{2 (\calE_{n0} - \calE_{00}) }{(\hbar\omega)^2 - (\calE_{n0}-\calE_{00})^2}I_n,
\label{ImsigmaIP}
\end{align}
where
 \begin{align}
I_n = -iN\left (  \la \bV_{00} |\calJ_+  | \bV_{n0}\ra \la \bV_{n0} |\calJ_+  | \bV_{00}\ra - c.c.\right )
\end{align}
Interestingly, the form of the complex Hall conductivity here is drastically different from that in the perpendicular magnetization phase, in that the real part is absorptive whereas the imaginary part is reactive. In particular, the dc Hall conductivity ${\rm Re }\,\sigma_H(0)$ vanishes completely.  

\begin{table}[htbp]
\centering
\[
\begin{array}{c|cccc} 
\tilde{D}_{1} & e & \tilde{s}_{1} & \tilde{s}^2_{1} & \tilde{s}^3_{1} \\
\hline 
B_{1} & 1 & 1 & 1 & 1 \\
B_{2} & 1 & -1  & 1 & -1 \\
B_{3} & 1 & i & -1 & -i\\
B_{4} & 1 & -i & -1 & i \\
E & 2 & 0 & 2 & 0 \\
\end{array}
\]
\caption{Character table of $\tilde D_1$}
\label{TabD1}
\end{table}

 \begin{figure}[htbp]
	\centering
	\includegraphics[width=8cm]{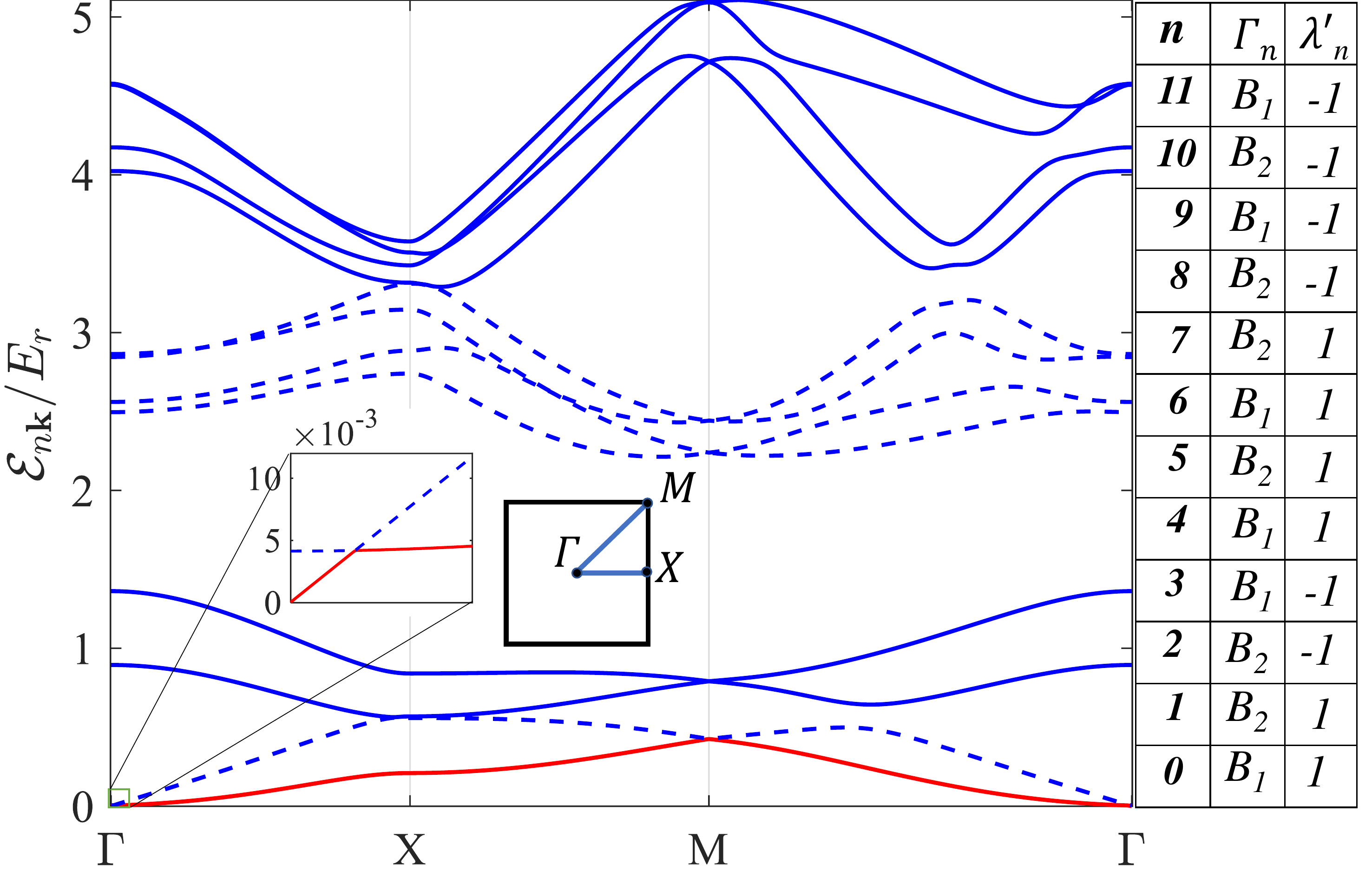}
	\caption{Bogliuobov bands in the in-plane magnetization phase. Although it appears that this phase has two gapless modes, the expanded view of the neighborhood of the $\Gamma$ point shows that this is not the case. On the right side, the $\Gamma_n$ and the $\lambda$ value corresponding to the $\bk = 0$ state of the $n$-th band are given. According to the selection rule  explained in the text, the transition is forbidden from the ground state to those bands indicated by dashed lines.} 
	\label{fig8}
\end{figure}

 \begin{figure}[htbp]
	\centering
	\includegraphics[width=8.5cm]{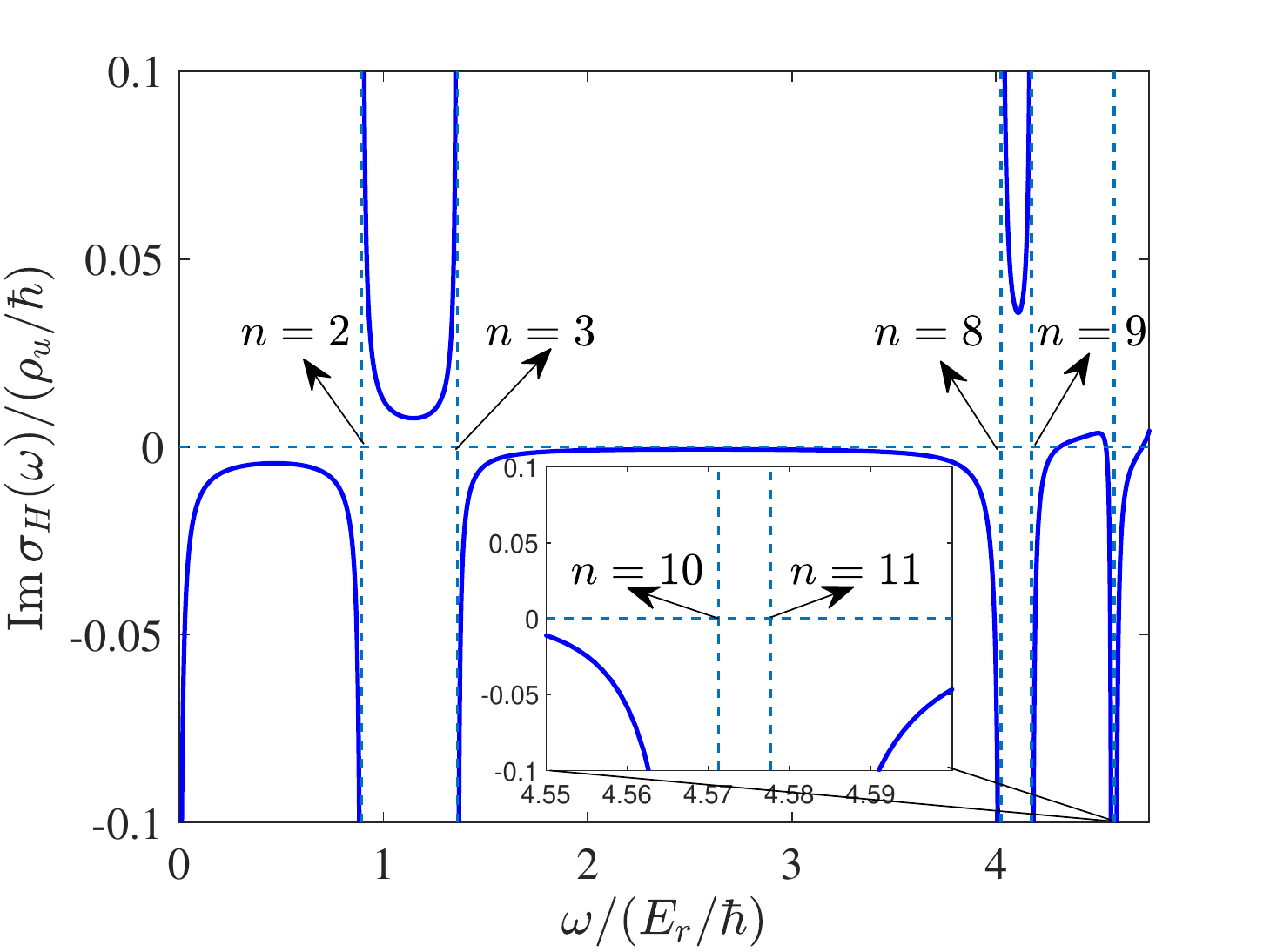}
	\caption{Imaginary part of the frequency-dependent Hall conductivity for the system in the in-plane magnetization phase. } 
	\label{fig9}
\end{figure}
Similar to previous analysis of the perpendicular magnetization phase, we can determine the selection rules of the matrix elements $\la \bV_{00} |\calJ_\pm  | \bV_{n0}\ra$ by first ascertaining the representations of the $\tilde D_1$ group whose basis functions correspond to  $\bV_{00}$, $\bV_{n0}$ and $\calJ_\pm$. For operators $\calJ_\pm$ that construct a higher than 1D representation, the generalization of the selection rule given earlier states that  the matrix elements $\la \bV_{00} |\calJ_\pm  | \bV_{n0}\ra$ vanish if $\Gamma_n$ is not contained in the 1D decomposition of the direct product $\Gamma_\calJ\otimes \Gamma_0$~\cite{Dresselhaus2007group}. From direct calculation and the character table, we find that $\Gamma_\calJ = E$ and thus
\begin{align}
 \Gamma_\calJ \otimes \Gamma_0 = B_1\oplus B_2. 
\end{align} 
As shown in Fig.~\ref{fig8}, it turns out  $\Gamma_n$ is in fact either $B_1$  or $B_2$ for the first $11$ excited bands, meaning that for these bands no restriction of transition is placed by this rule. However, this is not the case with respect to the selection rule of the nonsymmorphic symmetry. Recalling that the condensate here is invariant under $\Lambda_x\tilde r_2$, we can define the corresponding symmetry operation $f_2(\Lambda_x\tilde r_2)$ for the Bogoliubov Hamiltonian. Considering $I_n$ under the following transformations 
\begin{align}
f_2(\Lambda_x\tilde r_2) \calJ_{\pm }f^{-1}_2(\Lambda_x\tilde r_2) & = - \calJ_{\pm} \nn \\
f_2(\Lambda_x\tilde r_2) \bV_{n 0}  &= \lambda'_n \bV_{n0},
\end{align}
it is not difficult to see that $I_n\neq 0$ only if $\lambda'_n = -\lambda'_0$. According to the table of $\lambda'_n$ values shown in Fig.~\ref{fig8}, this rule then eliminates transitions to those states from the bands indicated by dashed lines. Again, this is consistent with the detailed calculations of $\rm Im\, \sigma_H(\omega)$ shown in Fig.~\ref{fig9}.


\section{dc Hall conductivity}
\label{DCHALL}
In the previous section we have demonstrated the existence of the anomalous Hall effect by explicit calculations of the frequency-dependent Hall conductivity. We found that the current transition matrix elements obey certain selection rules dictated by symmetry principals, which are used to explain the various resonances exhibited in the frequency-dependent Hall conductivities as well as the absence of the dc Hall conductivity in the in-plane magnetization phase.  However, these calculations do not provide underlying reasons for why the dc AHE is present in one phase but not in the other. In this section, we focus on the dc Hall conductivity and explore, from both a real and momentum space perspective, a deeper understanding of the causes of the AHE in the atomic superfluid. From the real space perspective, we argue that the chirality of the superfluid is directly responsible for a finite dc Hall conductivity; from the momentum space perspective, we show that finite Berry curvature at the condensation momentum of the noninteracting band underpins the dc Hall conductivity. 

\subsection{Real space perspective: chirality} 
In addition to the magnetization, chirality is another fundamental property that distinguishes the two magnetic phases. In contrast to the in-plane magnetization phase, the condensate in the perpendicular magnetization phase carries a finite total angular momentum and is thus a chiral superfluid. Such an atomic chiral superfluid is reminiscent of the fermionic chiral superfluid in $^3$He-A~\cite{2013Ikegami} or the chiral superconductor such as Sr$_2$RuO$_4$~\cite{2016Kallin}, although in these fermionic systems the angular momenta are carried by the relative motion of Cooper pairs.  Like these fermionic counterparts, the atomic chiral superfluid also gives rise to a nontrivial dichroic response~\cite{2017Tran,2020Midtgaard}. To be specific, let's consider the responses of the system to two rotating potentials
\begin{align}
V_{\pm}{(\br, t)} =2F (x\cos\omega t \pm y \sin\omega t),
\label{Vpm}
\end{align}
where $F$ is the force constant. 
 \begin{figure}[htbp]
	\centering
	\includegraphics[width=8.5cm]{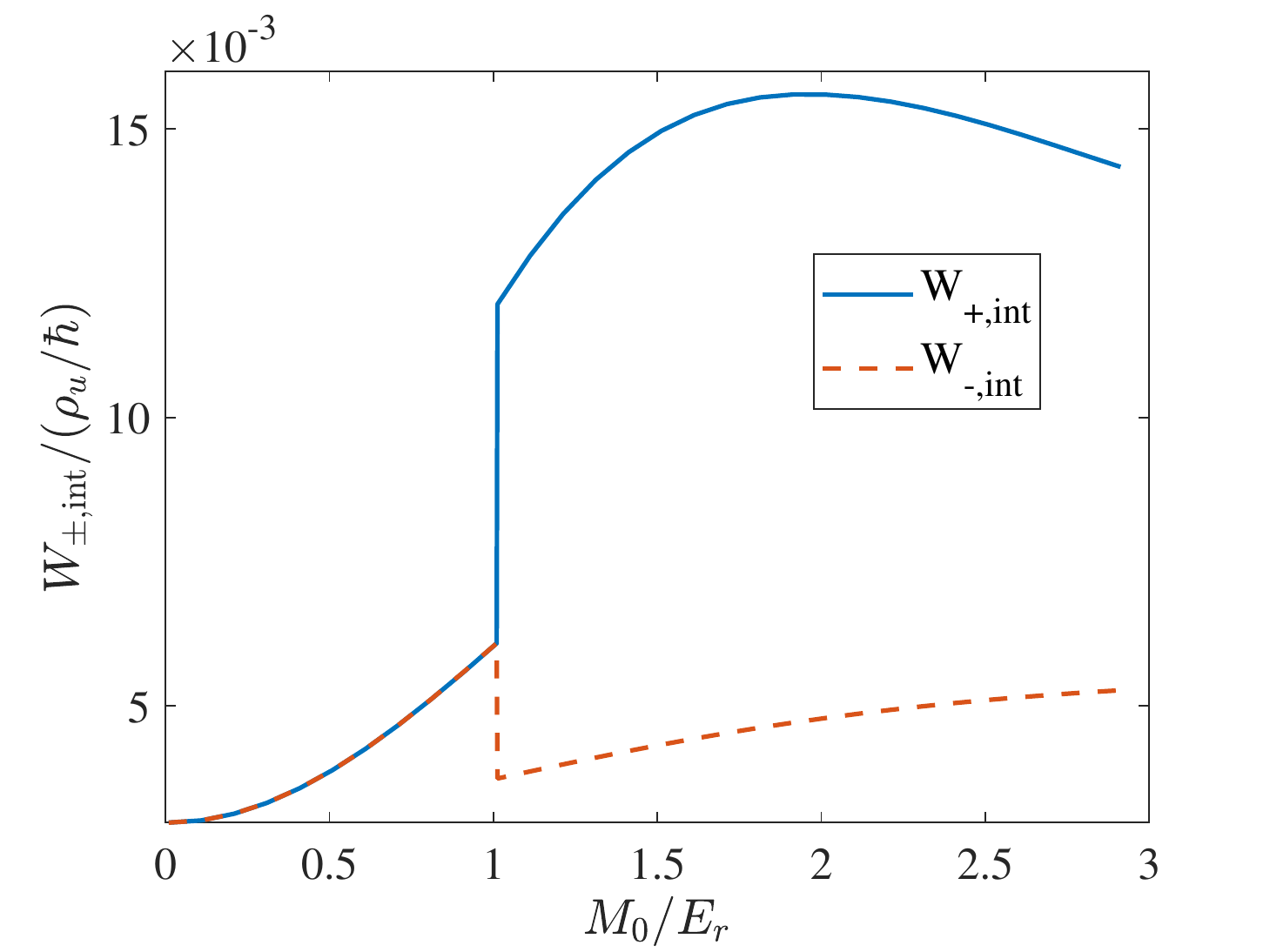}
	\caption{Integrated excitation rates $w_{\pm,\rm int}$ due to the rotating potentials in Eq.~(\ref{Vpm}),  as a function of the SO coupling strength.} 
	\label{fig10}
\end{figure}
Under such a perturbation,  the excitation rate out of the ground state per unit area can be calculated by the Fermi’s golden rule as~\cite{2017Tran,2020Midtgaard}
\begin{equation}
W_{\pm}(\omega)=\frac{2\pi}{\hbar \calA}\left(\frac{F}{\omega}\right)^2\sum_{n}\Big|\langle n|\hat J_\pm |0\rangle\Big|^2\delta(E_n-E_0-\hbar\omega).
\end{equation}
If the system is chiral, it naturally responds to the perturbations $V_+$ and $V_-$ differently such that $w_+(\omega) \neq w_-(\omega)$. 
Defining the following integrated excitation rates
\begin{align}
W_{\pm,\rm int} \equiv \frac{1}{4\pi F^2}\int_{0}^{\infty}\mathrm d\omega W_{\pm}(\omega),
\end{align}
 the difference of these two quantities in fact yields the real part of the dc Hall conductivity, namely~\cite{2017Tran,2020Midtgaard}
\begin{align}
{\rm Re}\,\sigma_H(0)={  W_{+,\rm int}-W_{-,\rm int}}. 
\end{align}
Thus, this analysis shows that the system's chirality implies a finite dc Hall conductivity. To demonstrate this explicitly, we show in Fig.~\ref{fig10} the calculated $W_{\pm,\rm int} $ as a function of the SO coupling strength, where we find  $W_{+,\rm int} = W_{-,\rm int} $ for the in-plane magnetization phase while  $W_{+,\rm int} > W_{-,\rm int} $ for the perpendicular magnetization phase. 

\subsection{Momentum space perspective: Berry curvature}
A complementary way of understanding the dc AHE in the atomic chiral superfluid is to make use of the relation between the dc Hall conductivity and the Berry curvature of the Bloch bands. In ferromagnetic materials where the anomalous Hall effect was originally discovered, the SO coupling gives rise to electronic bands with finite Berry curvatures, which can play the role of a magnetic field in deflecting the electrons moving under an electric field. The dc Hall conductivity is given by the summation of Berry curvatures of the occupied Bloch states.  Take the non-interacting bands in our SOC system as an example and suppose that fermions instead of bosons occupy these bands. The zero temperature dc Hall conductivity would then be given by 
\begin{align}
 {\rm Re}\, \sigma_H(0) = \frac{1}{\hbar\calA}\sum_{n\bk} \Omega_{n}(\bk),
 \end{align} 
 where the summation of $n,\bk$ is restricted to the occupied Bloch states. 
Here $\Omega_{n}(\bk)$ is the Berry curvature of the $n$-th band 
\begin{align}
\Omega_{n}(\bk) =i\left( \la \pa_{k_x}   \bm {\mathfrak {u}}_{n\bk}|\pa_{k_y}\bfu_{n\bk}\ra -\la \pa_{k_y} \bfu_{n\bk} |\pa_{k_x}\bfu_{n\bk}\ra \right ),
\end{align} 
where $\bfu_{n\bk}(\br) \equiv e^{-i\bk\cdot \br} \bphi_{n\bk}(\br)$ and $\la \pa_{k_x} \bfu_{n\bk} |\pa_{k_y}\bfu_{n\bk}\ra$ is a shorthand notation for the inner product of the spinor wave functions, $\sum_\sigma\int d\br \pa_{k_x}\mathfrak {u}_{n\bk\sigma}(\br) \pa_{k_y} \mathfrak {u}_{n\bk\sigma}(\br)$. Now, because of the double degeneracy of the bands, the Berry curvature $\Omega_n(\bk)$ is actually ambiguous without specifying the Bloch wave function $\bphi_{n\bk}(\br)$.  For spin polarized states $\bphi^\pm_{m\bk}$ introduced in Sec.~\ref{PT}, the Berry curvatures are generally finite. Let's consider the bottom two bands as an example. If we take $\bphi_{0,\bk} = \bphi^+_{0,\bk}$ and $\bphi_{1,\bk} = \bphi^-_{0,\bk}$, we find 
\begin{align}
\Omega_{0}(\bk) = - \Omega_{1}(\bk) \neq 0. 
\end{align}
On the other hand, for any state that is a superposition of 
$ \bphi^+_{m,\bk}$ and $ \bphi^-_{m,\bk}$ with equal weight, the Berry curvature is zero. 

 In our system, the origin of the dc Hall conductivity can also be discussed in this framework if we neglect the atomic interactions.
The dc Hall conductivity in Eq.~(\ref{ReS0}) can be written as
 \begin{align}
{\rm Re}\,  \sigma_H(0) & =-\frac{1}{\calA\hbar}{\rm Im}\sum_{n\neq 0}   \frac{J_{x,0n} J_{y,n0}-J_{y,0n} J_{x,n0}}{( E_{n0}- E_{00})^2} \nn\\
&=2\frac{N}{\calA\hbar}\sum_{n\neq 0}   \frac{|\calJ_{+,n0}|^2 - |\calJ_{- ,n0}|^2}{(\calE_{n0}-\calE_{00})^2},
\label{SigmaH}
\end{align}
where the second line is obtained by the Bogoliubov theory. In the non-interacting limit, the above expression reduces to
\begin{align}
{\rm Re}\,  \sigma_H(0) &= -2\frac{N}{\calA\hbar}\sum_{n\neq 0}   \frac{{\rm Im}\left [\la \bphi_{00}| p_x| \bphi_{n0}\ra \la \bphi_{n0}|p_y| \bphi_{00}\ra\right ] }{(\epsilon_{n0}-\epsilon_{00})^2} \nn \\
& = \frac{\rho_u}{\hbar S_u} \Omega_0 (\bk = 0),
\label{Snon}
\end{align}
where $S_u$ is the area of the Wigner-Seitz cell and $\rho_u$ is the number of atoms per cell. 
This is exactly what we expect because all the atoms condensate at the same $\bphi_{00}$ state in the non-interacting limit. Since $\Omega_0(\bk = 0)$ is finite in the perpendicular magnetization phase and is zero in the in-plane magnetization phase, the difference in the dc Hall conductivities between these two phases can be immediately understood from the above equation. In Fig.~\ref{fig10}, we have plotted  ${\rm Re}\,  \sigma_H(0) $ calculated  for various values of $(\rho g_{\uparrow\uparrow},\rho g_{\uparrow\downarrow}) = \zeta (0.35 E_r, 0.3 E_r)$, which clearly shows that the non-interacting result in Eq.~(\ref{Snon}) is recovered as $\zeta\rightarrow 0$.
 \begin{figure}[htbp]
	\centering
	\includegraphics[width=8.5cm]{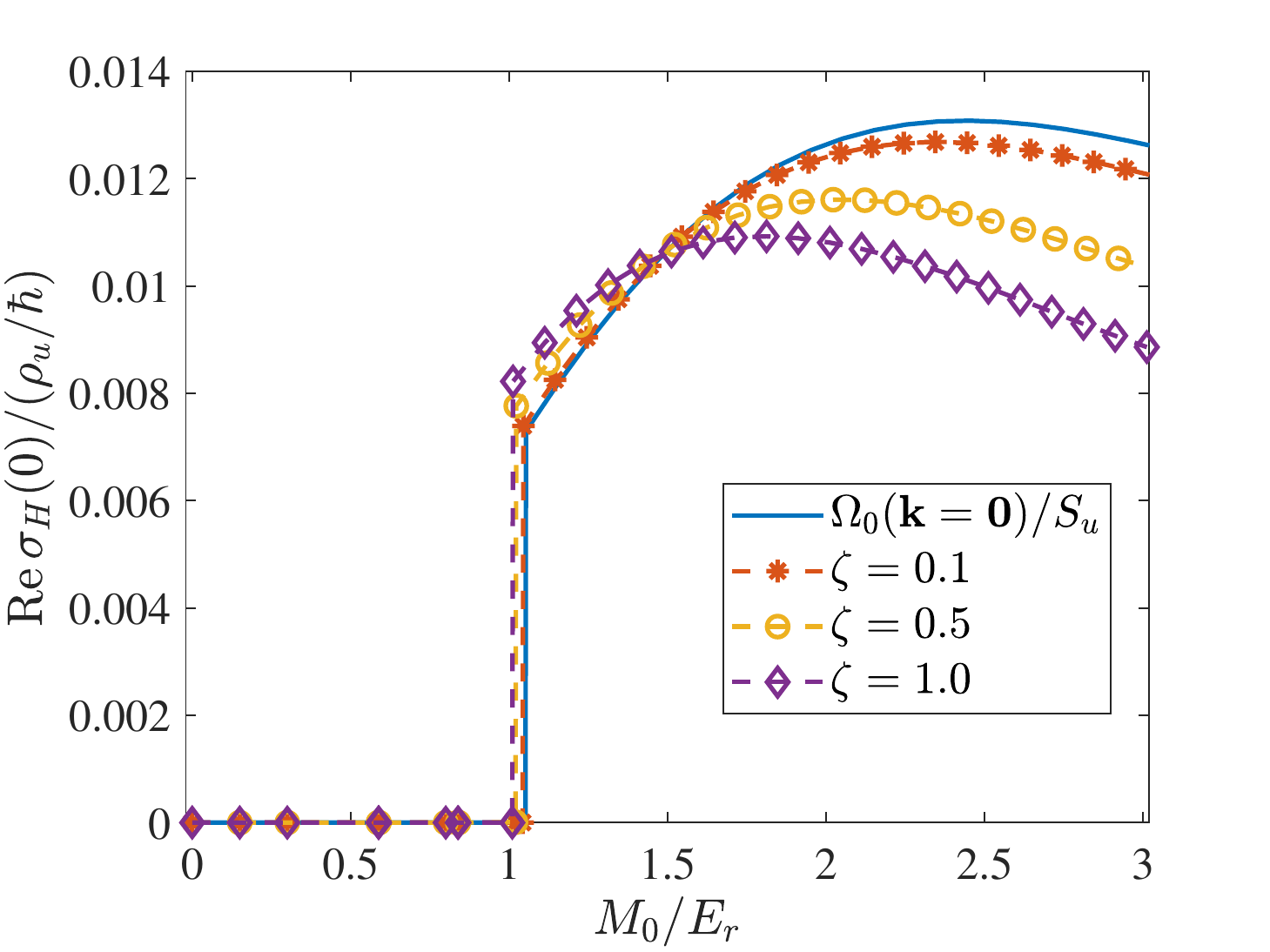}
	\caption{The dc Hall conductivity as a function of $M_0$ for various values of  interaction strengths. Here $(\rho g_{\uparrow\uparrow},\rho g_{\uparrow\downarrow}) = \zeta (0.35 E_r, 0.3 E_r)$. The solid line represents the right hand side of Eq.~(\ref{Snon}).}
	\label{fig10}
\end{figure}
\section{Experimental proposal}
\label{EP}
Since the SOC condensate has already been realized in experiments, it is crucial to ask whether the predicted AHE can be detected with currently available experimental tools. In trapped atomic systems it is difficult to directly measure the current response function  and hence the Hall conductivity. In earlier discussions on the relation between chirality and the anomalous Hall effect, we have in fact shown earlier that the dichroism probe can be used to measure the dc Hall conductivity. In this section we discuss a experimental method to probe the frequency dependence of the Hall conductivity and in particular to reveal the resonances exhibited in the Hall conductivity.  The idea is to make use of a close relation between the current and the center of mass (COM) response functions, so that one can deduce the conductivity tensor from the responses of the COM to a time-periodic linear potential.  This was first discussed in Ref.~\cite{2015Wu} and was later implemented experimentally to measure the longitudinal conductivity of a lattice Fermi gas~\cite{2019Anderson}. 

To be more specific, let's consider the transverse response of the center of mass (COM) degree of freedom, $\hat {\boldsymbol R} = \frac{1}{N}\sum_\sigma\int \br\hat \psi^\dag_\sigma(\br) \hat \psi_\sigma(\br )d\br $, to the following time-periodic linear potential 
\begin{equation}
	V(\br, t)=-Fx\cos\omega t, 
	\label{Vx}
\end{equation}
where $F$ is the force constant. If the system has a finite Hall conductivity, such a force along the $x$ direction will generate a response of the COM along the $y$ direction. Within linear response theory, the COM response can be written as 
\begin{align}
\langle \hat R_y(t)\rangle=R_y(\omega)\cos[\omega t-\phi_y(\omega)],
\end{align}
where $R_y(\omega)$ is the amplitude and $\phi_y(\omega)$ is the phase lagging. 
 It turns out that the finite-frequency Hall conductivity $\sigma_H(\omega)$ is related to the amplitude and phase of this response by~\cite{2015Wu} 
\begin{equation}
	\sigma_{H}(\omega)=\frac{N\omega}{i\mathcal A F }R_y(\omega)e^{i\phi_y(\omega)}.
	\label{Sigmacom}
\end{equation}	
Thus, by measuring the COM response along the $y$ direction to the potential in Eq.~(\ref{Vx}) we can infer the ac Hall conductivity. Now, from Eq.~(\ref{Sigmacom}) we may also calculate the expected behavior of the COM response from knowledge of $\sigma_H(\omega)$. Inverting Eq.~(\ref{Sigmacom}) we find
\begin{align}
R_y(\omega) & = \frac{\calA F}{N\omega} |\sigma_H(\omega)|  \\
\phi_y(\omega) & = \tan^{-1} \left (- {{\rm Re}\, \sigma(\omega)}/{{\rm Im} \,\sigma(\omega)}\right ).
\end{align}
In view of previous calculations of the Hall conductivity,  we expect to find resonant behavior for the amplitude of the responses in both phases.

\section{Concluding remarks}
\label{CR}
In this paper, we have calculated the finite frequency Hall conductivity of a 2D SOC Bose gas at zero temperature in the absence of an artificial magnetic field and demonstrated the existence of ground state intrinsic anomalous Hall effects across the magnetic phase transition of the system.  In the perpendicular magnetization phase, the SOC system realizes a chiral superfluid for which the real part of the ac Hall conductivity is reactive while the imaginary part is absorptive. Importantly we find a finite, albeit non-quantized, dc Hall conductivity in this phase. On the other hand, it is exactly the opposite in the in-plane magnetization phase, where the ac Hall conductivity has an absorptive real part and a reactive imaginary part, and the dc Hall conductivity vanishes. In both phases, the ac Hall responses exhibit various resonances in frequency which can be explained by the selection rules derived from the symmetry analysis. Furthermore, the contrast in the dc Hall conductivities of the two phases can be understood using the connection between the dc Hall conductivity and the chirality and the Berry curvature. Finally, we show that the AHEs discussed in this work can be readily probed experimentally by a measurement of COM responses. 

\section*{Acknowledgements}  This work is supported by National Key R$\&$D Program of China (Grant No. 2022YFA1404103), NSFC (Grant No.~11974161), Shenzhen Science and Technology Program (Grant No.~KQTD20200820113010023) and Key-Area Research and Development Program of Guangdong Province (Grant No.~2019B030330001).


\begin{thebibliography}{40}%
\makeatletter
\providecommand \@ifxundefined [1]{%
 \@ifx{#1\undefined}
}%
\providecommand \@ifnum [1]{%
 \ifnum #1\expandafter \@firstoftwo
 \else \expandafter \@secondoftwo
 \fi
}%
\providecommand \@ifx [1]{%
 \ifx #1\expandafter \@firstoftwo
 \else \expandafter \@secondoftwo
 \fi
}%
\providecommand \natexlab [1]{#1}%
\providecommand \enquote  [1]{``#1''}%
\providecommand \bibnamefont  [1]{#1}%
\providecommand \bibfnamefont [1]{#1}%
\providecommand \citenamefont [1]{#1}%
\providecommand \href@noop [0]{\@secondoftwo}%
\providecommand \href [0]{\begingroup \@sanitize@url \@href}%
\providecommand \@href[1]{\@@startlink{#1}\@@href}%
\providecommand \@@href[1]{\endgroup#1\@@endlink}%
\providecommand \@sanitize@url [0]{\catcode `\\12\catcode `\$12\catcode
  `\&12\catcode `\#12\catcode `\^12\catcode `\_12\catcode `\%12\relax}%
\providecommand \@@startlink[1]{}%
\providecommand \@@endlink[0]{}%
\providecommand \url  [0]{\begingroup\@sanitize@url \@url }%
\providecommand \@url [1]{\endgroup\@href {#1}{\urlprefix }}%
\providecommand \urlprefix  [0]{URL }%
\providecommand \Eprint [0]{\href }%
\providecommand \doibase [0]{http://dx.doi.org/}%
\providecommand \selectlanguage [0]{\@gobble}%
\providecommand \bibinfo  [0]{\@secondoftwo}%
\providecommand \bibfield  [0]{\@secondoftwo}%
\providecommand \translation [1]{[#1]}%
\providecommand \BibitemOpen [0]{}%
\providecommand \bibitemStop [0]{}%
\providecommand \bibitemNoStop [0]{.\EOS\space}%
\providecommand \EOS [0]{\spacefactor3000\relax}%
\providecommand \BibitemShut  [1]{\csname bibitem#1\endcsname}%
\let\auto@bib@innerbib\@empty
\bibitem [{\citenamefont {Hasan}\ and\ \citenamefont {Kane}(2010)}]{Hasan2010}%
  \BibitemOpen
  \bibfield  {author} {\bibinfo {author} {\bibfnamefont {M.~Z.}\ \bibnamefont
  {Hasan}}\ and\ \bibinfo {author} {\bibfnamefont {C.~L.}\ \bibnamefont
  {Kane}},\ }\href {\doibase 10.1103/RevModPhys.82.3045} {\bibfield  {journal}
  {\bibinfo  {journal} {Rev. Mod. Phys.}\ }\textbf {\bibinfo {volume} {82}},\
  \bibinfo {pages} {3045} (\bibinfo {year} {2010})}\BibitemShut {NoStop}%
\bibitem [{\citenamefont {Qi}\ and\ \citenamefont {Zhang}(2011)}]{Qi2011}%
  \BibitemOpen
  \bibfield  {author} {\bibinfo {author} {\bibfnamefont {X.-L.}\ \bibnamefont
  {Qi}}\ and\ \bibinfo {author} {\bibfnamefont {S.-C.}\ \bibnamefont {Zhang}},\
  }\href {\doibase 10.1103/RevModPhys.83.1057} {\bibfield  {journal} {\bibinfo
  {journal} {Rev. Mod. Phys.}\ }\textbf {\bibinfo {volume} {83}},\ \bibinfo
  {pages} {1057} (\bibinfo {year} {2011})}\BibitemShut {NoStop}%
\bibitem [{\citenamefont {Bernevig}\ and\ \citenamefont
  {Hughes}(2013)}]{2013Bernevig}%
  \BibitemOpen
  \bibfield  {author} {\bibinfo {author} {\bibfnamefont {B.~A.}\ \bibnamefont
  {Bernevig}}\ and\ \bibinfo {author} {\bibfnamefont {T.~L.}\ \bibnamefont
  {Hughes}},\ }\href@noop {} {\emph {\bibinfo {title} {Topological Insulators
  and Topological Superconductors}}}\ (\bibinfo  {publisher} {Princeton
  University Press, Princeton, NJ},\ \bibinfo {year} {2013})\BibitemShut
  {NoStop}%
\bibitem [{\citenamefont {Thouless}\ \emph {et~al.}(1982)\citenamefont
  {Thouless}, \citenamefont {Kohmoto}, \citenamefont {Nightingale},\ and\
  \citenamefont {den Nijs}}]{1982TKNN}%
  \BibitemOpen
  \bibfield  {author} {\bibinfo {author} {\bibfnamefont {D.~J.}\ \bibnamefont
  {Thouless}}, \bibinfo {author} {\bibfnamefont {M.}~\bibnamefont {Kohmoto}},
  \bibinfo {author} {\bibfnamefont {M.~P.}\ \bibnamefont {Nightingale}}, \ and\
  \bibinfo {author} {\bibfnamefont {M.}~\bibnamefont {den Nijs}},\ }\href
  {\doibase 10.1103/PhysRevLett.49.405} {\bibfield  {journal} {\bibinfo
  {journal} {Phys. Rev. Lett.}\ }\textbf {\bibinfo {volume} {49}},\ \bibinfo
  {pages} {405} (\bibinfo {year} {1982})}\BibitemShut {NoStop}%
\bibitem [{\citenamefont {Nagaosa}\ \emph {et~al.}(2010)\citenamefont
  {Nagaosa}, \citenamefont {Sinova}, \citenamefont {Onoda}, \citenamefont
  {MacDonald},\ and\ \citenamefont {Ong}}]{2010Nagaosa}%
  \BibitemOpen
  \bibfield  {author} {\bibinfo {author} {\bibfnamefont {N.}~\bibnamefont
  {Nagaosa}}, \bibinfo {author} {\bibfnamefont {J.}~\bibnamefont {Sinova}},
  \bibinfo {author} {\bibfnamefont {S.}~\bibnamefont {Onoda}}, \bibinfo
  {author} {\bibfnamefont {A.~H.}\ \bibnamefont {MacDonald}}, \ and\ \bibinfo
  {author} {\bibfnamefont {N.~P.}\ \bibnamefont {Ong}},\ }\href {\doibase
  10.1103/RevModPhys.82.1539} {\bibfield  {journal} {\bibinfo  {journal} {Rev.
  Mod. Phys.}\ }\textbf {\bibinfo {volume} {82}},\ \bibinfo {pages} {1539}
  (\bibinfo {year} {2010})}\BibitemShut {NoStop}%
\bibitem [{\citenamefont {Xiao}\ \emph {et~al.}(2010)\citenamefont {Xiao},
  \citenamefont {Chang},\ and\ \citenamefont {Niu}}]{2010Xiao}%
  \BibitemOpen
  \bibfield  {author} {\bibinfo {author} {\bibfnamefont {D.}~\bibnamefont
  {Xiao}}, \bibinfo {author} {\bibfnamefont {M.-C.}\ \bibnamefont {Chang}}, \
  and\ \bibinfo {author} {\bibfnamefont {Q.}~\bibnamefont {Niu}},\ }\href
  {\doibase 10.1103/RevModPhys.82.1959} {\bibfield  {journal} {\bibinfo
  {journal} {Rev. Mod. Phys.}\ }\textbf {\bibinfo {volume} {82}},\ \bibinfo
  {pages} {1959} (\bibinfo {year} {2010})}\BibitemShut {NoStop}%
\bibitem [{\citenamefont {Cooper}\ \emph {et~al.}(2019)\citenamefont {Cooper},
  \citenamefont {Dalibard},\ and\ \citenamefont {Spielman}}]{Cooper2019}%
  \BibitemOpen
  \bibfield  {author} {\bibinfo {author} {\bibfnamefont {N.~R.}\ \bibnamefont
  {Cooper}}, \bibinfo {author} {\bibfnamefont {J.}~\bibnamefont {Dalibard}}, \
  and\ \bibinfo {author} {\bibfnamefont {I.~B.}\ \bibnamefont {Spielman}},\
  }\href {\doibase 10.1103/RevModPhys.91.015005} {\bibfield  {journal}
  {\bibinfo  {journal} {Rev. Mod. Phys.}\ }\textbf {\bibinfo {volume} {91}},\
  \bibinfo {pages} {015005} (\bibinfo {year} {2019})}\BibitemShut {NoStop}%
\bibitem [{\citenamefont {Aidelsburger}\ \emph {et~al.}(2013)\citenamefont
  {Aidelsburger}, \citenamefont {Atala}, \citenamefont {Lohse}, \citenamefont
  {Barreiro}, \citenamefont {Paredes},\ and\ \citenamefont
  {Bloch}}]{Aidelsburger2013}%
  \BibitemOpen
  \bibfield  {author} {\bibinfo {author} {\bibfnamefont {M.}~\bibnamefont
  {Aidelsburger}}, \bibinfo {author} {\bibfnamefont {M.}~\bibnamefont {Atala}},
  \bibinfo {author} {\bibfnamefont {M.}~\bibnamefont {Lohse}}, \bibinfo
  {author} {\bibfnamefont {J.~T.}\ \bibnamefont {Barreiro}}, \bibinfo {author}
  {\bibfnamefont {B.}~\bibnamefont {Paredes}}, \ and\ \bibinfo {author}
  {\bibfnamefont {I.}~\bibnamefont {Bloch}},\ }\href {\doibase
  10.1103/PhysRevLett.111.185301} {\bibfield  {journal} {\bibinfo  {journal}
  {Phys. Rev. Lett.}\ }\textbf {\bibinfo {volume} {111}},\ \bibinfo {pages}
  {185301} (\bibinfo {year} {2013})}\BibitemShut {NoStop}%
\bibitem [{\citenamefont {Miyake}\ \emph {et~al.}(2013)\citenamefont {Miyake},
  \citenamefont {Siviloglou}, \citenamefont {Kennedy}, \citenamefont {Burton},\
  and\ \citenamefont {Ketterle}}]{Miyake2013}%
  \BibitemOpen
  \bibfield  {author} {\bibinfo {author} {\bibfnamefont {H.}~\bibnamefont
  {Miyake}}, \bibinfo {author} {\bibfnamefont {G.~A.}\ \bibnamefont
  {Siviloglou}}, \bibinfo {author} {\bibfnamefont {C.~J.}\ \bibnamefont
  {Kennedy}}, \bibinfo {author} {\bibfnamefont {W.~C.}\ \bibnamefont {Burton}},
  \ and\ \bibinfo {author} {\bibfnamefont {W.}~\bibnamefont {Ketterle}},\
  }\href {\doibase 10.1103/PhysRevLett.111.185302} {\bibfield  {journal}
  {\bibinfo  {journal} {Phys. Rev. Lett.}\ }\textbf {\bibinfo {volume} {111}},\
  \bibinfo {pages} {185302} (\bibinfo {year} {2013})}\BibitemShut {NoStop}%
\bibitem [{\citenamefont {Atala}\ \emph {et~al.}(2013)\citenamefont {Atala},
  \citenamefont {Aidelsburger}, \citenamefont {Barreiro}, \citenamefont
  {Abanin}, \citenamefont {Kitagawa}, \citenamefont {Demler},\ and\
  \citenamefont {Bloch}}]{atala2013}%
  \BibitemOpen
  \bibfield  {author} {\bibinfo {author} {\bibfnamefont {M.}~\bibnamefont
  {Atala}}, \bibinfo {author} {\bibfnamefont {M.}~\bibnamefont {Aidelsburger}},
  \bibinfo {author} {\bibfnamefont {J.~T.}\ \bibnamefont {Barreiro}}, \bibinfo
  {author} {\bibfnamefont {D.}~\bibnamefont {Abanin}}, \bibinfo {author}
  {\bibfnamefont {T.}~\bibnamefont {Kitagawa}}, \bibinfo {author}
  {\bibfnamefont {E.}~\bibnamefont {Demler}}, \ and\ \bibinfo {author}
  {\bibfnamefont {I.}~\bibnamefont {Bloch}},\ }\href {\doibase
  10.1038/nphys2790} {\bibfield  {journal} {\bibinfo  {journal} {Nature
  Physics}\ }\textbf {\bibinfo {volume} {9}},\ \bibinfo {pages} {795} (\bibinfo
  {year} {2013})}\BibitemShut {NoStop}%
\bibitem [{\citenamefont {Jotzu}\ \emph {et~al.}(2014)\citenamefont {Jotzu},
  \citenamefont {Messer}, \citenamefont {Desbuquois}, \citenamefont {Lebrat},
  \citenamefont {Uehlinger}, \citenamefont {Greif},\ and\ \citenamefont
  {Esslinger}}]{Jotzu2014}%
  \BibitemOpen
  \bibfield  {author} {\bibinfo {author} {\bibfnamefont {G.}~\bibnamefont
  {Jotzu}}, \bibinfo {author} {\bibfnamefont {M.}~\bibnamefont {Messer}},
  \bibinfo {author} {\bibfnamefont {R.}~\bibnamefont {Desbuquois}}, \bibinfo
  {author} {\bibfnamefont {M.}~\bibnamefont {Lebrat}}, \bibinfo {author}
  {\bibfnamefont {T.}~\bibnamefont {Uehlinger}}, \bibinfo {author}
  {\bibfnamefont {D.}~\bibnamefont {Greif}}, \ and\ \bibinfo {author}
  {\bibfnamefont {T.}~\bibnamefont {Esslinger}},\ }\href {\doibase
  10.1038/nature13915} {\bibfield  {journal} {\bibinfo  {journal} {Nature}\
  }\textbf {\bibinfo {volume} {515}},\ \bibinfo {pages} {237} (\bibinfo {year}
  {2014})}\BibitemShut {NoStop}%
\bibitem [{\citenamefont {Xu}\ \emph {et~al.}(2016)\citenamefont {Xu},
  \citenamefont {You}, \citenamefont {Hemmerich},\ and\ \citenamefont
  {Liu}}]{Xu2016}%
  \BibitemOpen
  \bibfield  {author} {\bibinfo {author} {\bibfnamefont {Z.-F.}\ \bibnamefont
  {Xu}}, \bibinfo {author} {\bibfnamefont {L.}~\bibnamefont {You}}, \bibinfo
  {author} {\bibfnamefont {A.}~\bibnamefont {Hemmerich}}, \ and\ \bibinfo
  {author} {\bibfnamefont {W.~V.}\ \bibnamefont {Liu}},\ }\href {\doibase
  10.1103/PhysRevLett.117.085301} {\bibfield  {journal} {\bibinfo  {journal}
  {Phys. Rev. Lett.}\ }\textbf {\bibinfo {volume} {117}},\ \bibinfo {pages}
  {085301} (\bibinfo {year} {2016})}\BibitemShut {NoStop}%
\bibitem [{\citenamefont {Kock}\ \emph {et~al.}(2016)\citenamefont {Kock},
  \citenamefont {Hippler}, \citenamefont {Ewerbeck},\ and\ \citenamefont
  {Hemmerich}}]{2016Kock}%
  \BibitemOpen
  \bibfield  {author} {\bibinfo {author} {\bibfnamefont {T.}~\bibnamefont
  {Kock}}, \bibinfo {author} {\bibfnamefont {C.}~\bibnamefont {Hippler}},
  \bibinfo {author} {\bibfnamefont {A.}~\bibnamefont {Ewerbeck}}, \ and\
  \bibinfo {author} {\bibfnamefont {A.}~\bibnamefont {Hemmerich}},\ }\href
  {\doibase 10.1088/0953-4075/49/4/042001} {\bibfield  {journal} {\bibinfo
  {journal} {Journal of Physics B: Atomic, Molecular and Optical Physics}\
  }\textbf {\bibinfo {volume} {49}},\ \bibinfo {pages} {042001} (\bibinfo
  {year} {2016})}\BibitemShut {NoStop}%
\bibitem [{\citenamefont {Li}\ and\ \citenamefont
  {Liu}(2016)}]{2016XiaopengLi}%
  \BibitemOpen
  \bibfield  {author} {\bibinfo {author} {\bibfnamefont {X.}~\bibnamefont
  {Li}}\ and\ \bibinfo {author} {\bibfnamefont {W.~V.}\ \bibnamefont {Liu}},\
  }\href {\doibase 10.1088/0034-4885/79/11/116401} {\bibfield  {journal}
  {\bibinfo  {journal} {Reports on Progress in Physics}\ }\textbf {\bibinfo
  {volume} {79}},\ \bibinfo {pages} {116401} (\bibinfo {year}
  {2016})}\BibitemShut {NoStop}%
\bibitem [{\citenamefont {Di~Liberto}\ \emph {et~al.}(2016)\citenamefont
  {Di~Liberto}, \citenamefont {Hemmerich},\ and\ \citenamefont
  {Morais~Smith}}]{2016Liberto}%
  \BibitemOpen
  \bibfield  {author} {\bibinfo {author} {\bibfnamefont {M.}~\bibnamefont
  {Di~Liberto}}, \bibinfo {author} {\bibfnamefont {A.}~\bibnamefont
  {Hemmerich}}, \ and\ \bibinfo {author} {\bibfnamefont {C.}~\bibnamefont
  {Morais~Smith}},\ }\href {\doibase 10.1103/PhysRevLett.117.163001} {\bibfield
   {journal} {\bibinfo  {journal} {Phys. Rev. Lett.}\ }\textbf {\bibinfo
  {volume} {117}},\ \bibinfo {pages} {163001} (\bibinfo {year}
  {2016})}\BibitemShut {NoStop}%
\bibitem [{\citenamefont {M\"uller}\ \emph {et~al.}(2007)\citenamefont
  {M\"uller}, \citenamefont {F\"olling}, \citenamefont {Widera},\ and\
  \citenamefont {Bloch}}]{2007Muller}%
  \BibitemOpen
  \bibfield  {author} {\bibinfo {author} {\bibfnamefont {T.}~\bibnamefont
  {M\"uller}}, \bibinfo {author} {\bibfnamefont {S.}~\bibnamefont {F\"olling}},
  \bibinfo {author} {\bibfnamefont {A.}~\bibnamefont {Widera}}, \ and\ \bibinfo
  {author} {\bibfnamefont {I.}~\bibnamefont {Bloch}},\ }\href {\doibase
  10.1103/PhysRevLett.99.200405} {\bibfield  {journal} {\bibinfo  {journal}
  {Phys. Rev. Lett.}\ }\textbf {\bibinfo {volume} {99}},\ \bibinfo {pages}
  {200405} (\bibinfo {year} {2007})}\BibitemShut {NoStop}%
\bibitem [{\citenamefont {Wirth}\ \emph {et~al.}(2011)\citenamefont {Wirth},
  \citenamefont {{\"O}lschl{\"a}ger},\ and\ \citenamefont
  {Hemmerich}}]{2010Wirth}%
  \BibitemOpen
  \bibfield  {author} {\bibinfo {author} {\bibfnamefont {G.}~\bibnamefont
  {Wirth}}, \bibinfo {author} {\bibfnamefont {M.}~\bibnamefont
  {{\"O}lschl{\"a}ger}}, \ and\ \bibinfo {author} {\bibfnamefont
  {A.}~\bibnamefont {Hemmerich}},\ }\href {\doibase 10.1038/nphys1857}
  {\bibfield  {journal} {\bibinfo  {journal} {Nature Physics}\ }\textbf
  {\bibinfo {volume} {7}},\ \bibinfo {pages} {147} (\bibinfo {year}
  {2011})}\BibitemShut {NoStop}%
\bibitem [{\citenamefont {Soltan-Panahi}\ \emph {et~al.}(2012)\citenamefont
  {Soltan-Panahi}, \citenamefont {L{\"u}hmann}, \citenamefont {Struck},
  \citenamefont {Windpassinger},\ and\ \citenamefont {Sengstock}}]{2012Soltan}%
  \BibitemOpen
  \bibfield  {author} {\bibinfo {author} {\bibfnamefont {P.}~\bibnamefont
  {Soltan-Panahi}}, \bibinfo {author} {\bibfnamefont {D.-S.}\ \bibnamefont
  {L{\"u}hmann}}, \bibinfo {author} {\bibfnamefont {J.}~\bibnamefont {Struck}},
  \bibinfo {author} {\bibfnamefont {P.}~\bibnamefont {Windpassinger}}, \ and\
  \bibinfo {author} {\bibfnamefont {K.}~\bibnamefont {Sengstock}},\ }\href
  {\doibase 10.1038/nphys2128} {\bibfield  {journal} {\bibinfo  {journal}
  {Nature Physics}\ }\textbf {\bibinfo {volume} {8}},\ \bibinfo {pages} {71}
  (\bibinfo {year} {2012})}\BibitemShut {NoStop}%
\bibitem [{\citenamefont {Ölschläger}\ \emph {et~al.}(2013)\citenamefont
  {Ölschläger}, \citenamefont {Kock}, \citenamefont {Wirth}, \citenamefont
  {Ewerbeck}, \citenamefont {Smith},\ and\ \citenamefont
  {Hemmerich}}]{2013Olschlager}%
  \BibitemOpen
  \bibfield  {author} {\bibinfo {author} {\bibfnamefont {M.}~\bibnamefont
  {Ölschläger}}, \bibinfo {author} {\bibfnamefont {T.}~\bibnamefont {Kock}},
  \bibinfo {author} {\bibfnamefont {G.}~\bibnamefont {Wirth}}, \bibinfo
  {author} {\bibfnamefont {A.}~\bibnamefont {Ewerbeck}}, \bibinfo {author}
  {\bibfnamefont {C.~M.}\ \bibnamefont {Smith}}, \ and\ \bibinfo {author}
  {\bibfnamefont {A.}~\bibnamefont {Hemmerich}},\ }\href {\doibase
  10.1088/1367-2630/15/8/083041} {\bibfield  {journal} {\bibinfo  {journal}
  {New Journal of Physics}\ }\textbf {\bibinfo {volume} {15}},\ \bibinfo
  {pages} {083041} (\bibinfo {year} {2013})}\BibitemShut {NoStop}%
\bibitem [{\citenamefont {Kock}\ \emph {et~al.}(2015)\citenamefont {Kock},
  \citenamefont {\"Olschl\"ager}, \citenamefont {Ewerbeck}, \citenamefont
  {Huang}, \citenamefont {Mathey},\ and\ \citenamefont {Hemmerich}}]{2015Kock}%
  \BibitemOpen
  \bibfield  {author} {\bibinfo {author} {\bibfnamefont {T.}~\bibnamefont
  {Kock}}, \bibinfo {author} {\bibfnamefont {M.}~\bibnamefont
  {\"Olschl\"ager}}, \bibinfo {author} {\bibfnamefont {A.}~\bibnamefont
  {Ewerbeck}}, \bibinfo {author} {\bibfnamefont {W.-M.}\ \bibnamefont {Huang}},
  \bibinfo {author} {\bibfnamefont {L.}~\bibnamefont {Mathey}}, \ and\ \bibinfo
  {author} {\bibfnamefont {A.}~\bibnamefont {Hemmerich}},\ }\href {\doibase
  10.1103/PhysRevLett.114.115301} {\bibfield  {journal} {\bibinfo  {journal}
  {Phys. Rev. Lett.}\ }\textbf {\bibinfo {volume} {114}},\ \bibinfo {pages}
  {115301} (\bibinfo {year} {2015})}\BibitemShut {NoStop}%
\bibitem [{\citenamefont {Weinberg}\ \emph {et~al.}(2016)\citenamefont
  {Weinberg}, \citenamefont {Staarmann}, \citenamefont {Ölschläger},
  \citenamefont {Simonet},\ and\ \citenamefont {Sengstock}}]{2016Sengstock}%
  \BibitemOpen
  \bibfield  {author} {\bibinfo {author} {\bibfnamefont {M.}~\bibnamefont
  {Weinberg}}, \bibinfo {author} {\bibfnamefont {C.}~\bibnamefont {Staarmann}},
  \bibinfo {author} {\bibfnamefont {C.}~\bibnamefont {Ölschläger}}, \bibinfo
  {author} {\bibfnamefont {J.}~\bibnamefont {Simonet}}, \ and\ \bibinfo
  {author} {\bibfnamefont {K.}~\bibnamefont {Sengstock}},\ }\href {\doibase
  10.1088/2053-1583/3/2/024005} {\bibfield  {journal} {\bibinfo  {journal} {2D
  Materials}\ }\textbf {\bibinfo {volume} {3}},\ \bibinfo {pages} {024005}
  (\bibinfo {year} {2016})}\BibitemShut {NoStop}%
\bibitem [{\citenamefont {Sun}\ \emph {et~al.}(2018)\citenamefont {Sun},
  \citenamefont {Wang}, \citenamefont {Xu}, \citenamefont {Yi}, \citenamefont
  {Zhang}, \citenamefont {Wu}, \citenamefont {Deng}, \citenamefont {Liu},
  \citenamefont {Chen},\ and\ \citenamefont {Pan}}]{Sun2018}%
  \BibitemOpen
  \bibfield  {author} {\bibinfo {author} {\bibfnamefont {W.}~\bibnamefont
  {Sun}}, \bibinfo {author} {\bibfnamefont {B.-Z.}\ \bibnamefont {Wang}},
  \bibinfo {author} {\bibfnamefont {X.-T.}\ \bibnamefont {Xu}}, \bibinfo
  {author} {\bibfnamefont {C.-R.}\ \bibnamefont {Yi}}, \bibinfo {author}
  {\bibfnamefont {L.}~\bibnamefont {Zhang}}, \bibinfo {author} {\bibfnamefont
  {Z.}~\bibnamefont {Wu}}, \bibinfo {author} {\bibfnamefont {Y.}~\bibnamefont
  {Deng}}, \bibinfo {author} {\bibfnamefont {X.-J.}\ \bibnamefont {Liu}},
  \bibinfo {author} {\bibfnamefont {S.}~\bibnamefont {Chen}}, \ and\ \bibinfo
  {author} {\bibfnamefont {J.-W.}\ \bibnamefont {Pan}},\ }\href {\doibase
  10.1103/PhysRevLett.121.150401} {\bibfield  {journal} {\bibinfo  {journal}
  {Phys. Rev. Lett.}\ }\textbf {\bibinfo {volume} {121}},\ \bibinfo {pages}
  {150401} (\bibinfo {year} {2018})}\BibitemShut {NoStop}%
\bibitem [{\citenamefont {Hachmann}\ \emph {et~al.}(2021)\citenamefont
  {Hachmann}, \citenamefont {Kiefer}, \citenamefont {Riebesehl}, \citenamefont
  {Eichberger},\ and\ \citenamefont {Hemmerich}}]{2021Hachmann}%
  \BibitemOpen
  \bibfield  {author} {\bibinfo {author} {\bibfnamefont {M.}~\bibnamefont
  {Hachmann}}, \bibinfo {author} {\bibfnamefont {Y.}~\bibnamefont {Kiefer}},
  \bibinfo {author} {\bibfnamefont {J.}~\bibnamefont {Riebesehl}}, \bibinfo
  {author} {\bibfnamefont {R.}~\bibnamefont {Eichberger}}, \ and\ \bibinfo
  {author} {\bibfnamefont {A.}~\bibnamefont {Hemmerich}},\ }\href {\doibase
  10.1103/PhysRevLett.127.033201} {\bibfield  {journal} {\bibinfo  {journal}
  {Phys. Rev. Lett.}\ }\textbf {\bibinfo {volume} {127}},\ \bibinfo {pages}
  {033201} (\bibinfo {year} {2021})}\BibitemShut {NoStop}%
\bibitem [{\citenamefont {Vargas}\ \emph {et~al.}(2021)\citenamefont {Vargas},
  \citenamefont {Nuske}, \citenamefont {Eichberger}, \citenamefont {Hippler},
  \citenamefont {Mathey},\ and\ \citenamefont {Hemmerich}}]{2021Vargas}%
  \BibitemOpen
  \bibfield  {author} {\bibinfo {author} {\bibfnamefont {J.}~\bibnamefont
  {Vargas}}, \bibinfo {author} {\bibfnamefont {M.}~\bibnamefont {Nuske}},
  \bibinfo {author} {\bibfnamefont {R.}~\bibnamefont {Eichberger}}, \bibinfo
  {author} {\bibfnamefont {C.}~\bibnamefont {Hippler}}, \bibinfo {author}
  {\bibfnamefont {L.}~\bibnamefont {Mathey}}, \ and\ \bibinfo {author}
  {\bibfnamefont {A.}~\bibnamefont {Hemmerich}},\ }\href {\doibase
  10.1103/PhysRevLett.126.200402} {\bibfield  {journal} {\bibinfo  {journal}
  {Phys. Rev. Lett.}\ }\textbf {\bibinfo {volume} {126}},\ \bibinfo {pages}
  {200402} (\bibinfo {year} {2021})}\BibitemShut {NoStop}%
\bibitem [{\citenamefont {Jin}\ \emph {et~al.}(2021)\citenamefont {Jin},
  \citenamefont {Zhang}, \citenamefont {Guo}, \citenamefont {Chen},
  \citenamefont {Zhou},\ and\ \citenamefont {Li}}]{2021Jin}%
  \BibitemOpen
  \bibfield  {author} {\bibinfo {author} {\bibfnamefont {S.}~\bibnamefont
  {Jin}}, \bibinfo {author} {\bibfnamefont {W.}~\bibnamefont {Zhang}}, \bibinfo
  {author} {\bibfnamefont {X.}~\bibnamefont {Guo}}, \bibinfo {author}
  {\bibfnamefont {X.}~\bibnamefont {Chen}}, \bibinfo {author} {\bibfnamefont
  {X.}~\bibnamefont {Zhou}}, \ and\ \bibinfo {author} {\bibfnamefont
  {X.}~\bibnamefont {Li}},\ }\href {\doibase 10.1103/PhysRevLett.126.035301}
  {\bibfield  {journal} {\bibinfo  {journal} {Phys. Rev. Lett.}\ }\textbf
  {\bibinfo {volume} {126}},\ \bibinfo {pages} {035301} (\bibinfo {year}
  {2021})}\BibitemShut {NoStop}%
\bibitem [{\citenamefont {Wang}\ \emph {et~al.}(2021)\citenamefont {Wang},
  \citenamefont {Luo}, \citenamefont {Liu}, \citenamefont {Liu}, \citenamefont
  {Hemmerich},\ and\ \citenamefont {Xu}}]{2021Wang}%
  \BibitemOpen
  \bibfield  {author} {\bibinfo {author} {\bibfnamefont {X.-Q.}\ \bibnamefont
  {Wang}}, \bibinfo {author} {\bibfnamefont {G.-Q.}\ \bibnamefont {Luo}},
  \bibinfo {author} {\bibfnamefont {J.-Y.}\ \bibnamefont {Liu}}, \bibinfo
  {author} {\bibfnamefont {W.~V.}\ \bibnamefont {Liu}}, \bibinfo {author}
  {\bibfnamefont {A.}~\bibnamefont {Hemmerich}}, \ and\ \bibinfo {author}
  {\bibfnamefont {Z.-F.}\ \bibnamefont {Xu}},\ }\href {\doibase
  10.1038/s41586-021-03702-0} {\bibfield  {journal} {\bibinfo  {journal}
  {Nature}\ }\textbf {\bibinfo {volume} {596}},\ \bibinfo {pages} {227}
  (\bibinfo {year} {2021})}\BibitemShut {NoStop}%
\bibitem [{\citenamefont {Chien}\ \emph {et~al.}(2015)\citenamefont {Chien},
  \citenamefont {Peotta},\ and\ \citenamefont {Di~Ventra}}]{Chien2015}%
  \BibitemOpen
  \bibfield  {author} {\bibinfo {author} {\bibfnamefont {C.-C.}\ \bibnamefont
  {Chien}}, \bibinfo {author} {\bibfnamefont {S.}~\bibnamefont {Peotta}}, \
  and\ \bibinfo {author} {\bibfnamefont {M.}~\bibnamefont {Di~Ventra}},\ }\href
  {\doibase 10.1038/nphys3531} {\bibfield  {journal} {\bibinfo  {journal}
  {Nature Physics}\ }\textbf {\bibinfo {volume} {11}},\ \bibinfo {pages} {998}
  (\bibinfo {year} {2015})}\BibitemShut {NoStop}%
\bibitem [{\citenamefont {Wu}\ \emph {et~al.}(2015)\citenamefont {Wu},
  \citenamefont {Taylor},\ and\ \citenamefont {Zaremba}}]{2015Wu}%
  \BibitemOpen
  \bibfield  {author} {\bibinfo {author} {\bibfnamefont {Z.}~\bibnamefont
  {Wu}}, \bibinfo {author} {\bibfnamefont {E.}~\bibnamefont {Taylor}}, \ and\
  \bibinfo {author} {\bibfnamefont {E.}~\bibnamefont {Zaremba}},\ }\href
  {\doibase 10.1209/0295-5075/110/26002} {\bibfield  {journal} {\bibinfo
  {journal} {{EPL} (Europhysics Letters)}\ }\textbf {\bibinfo {volume} {110}},\
  \bibinfo {pages} {26002} (\bibinfo {year} {2015})}\BibitemShut {NoStop}%
\bibitem [{\citenamefont {Anderson}\ \emph {et~al.}(2019)\citenamefont
  {Anderson}, \citenamefont {Wang}, \citenamefont {Xu}, \citenamefont {Venu},
  \citenamefont {Trotzky}, \citenamefont {Chevy},\ and\ \citenamefont
  {Thywissen}}]{2019Anderson}%
  \BibitemOpen
  \bibfield  {author} {\bibinfo {author} {\bibfnamefont {R.}~\bibnamefont
  {Anderson}}, \bibinfo {author} {\bibfnamefont {F.}~\bibnamefont {Wang}},
  \bibinfo {author} {\bibfnamefont {P.}~\bibnamefont {Xu}}, \bibinfo {author}
  {\bibfnamefont {V.}~\bibnamefont {Venu}}, \bibinfo {author} {\bibfnamefont
  {S.}~\bibnamefont {Trotzky}}, \bibinfo {author} {\bibfnamefont
  {F.}~\bibnamefont {Chevy}}, \ and\ \bibinfo {author} {\bibfnamefont {J.~H.}\
  \bibnamefont {Thywissen}},\ }\href {\doibase 10.1103/PhysRevLett.122.153602}
  {\bibfield  {journal} {\bibinfo  {journal} {Phys. Rev. Lett.}\ }\textbf
  {\bibinfo {volume} {122}},\ \bibinfo {pages} {153602} (\bibinfo {year}
  {2019})}\BibitemShut {NoStop}%
\bibitem [{\citenamefont {Huang}\ \emph {et~al.}(2022)\citenamefont {Huang},
  \citenamefont {Xu},\ and\ \citenamefont {Wu}}]{Huang2022}%
  \BibitemOpen
  \bibfield  {author} {\bibinfo {author} {\bibfnamefont {G.-H.}\ \bibnamefont
  {Huang}}, \bibinfo {author} {\bibfnamefont {Z.-F.}\ \bibnamefont {Xu}}, \
  and\ \bibinfo {author} {\bibfnamefont {Z.}~\bibnamefont {Wu}},\ }\href
  {\doibase 10.1103/PhysRevLett.129.185301} {\bibfield  {journal} {\bibinfo
  {journal} {Phys. Rev. Lett.}\ }\textbf {\bibinfo {volume} {129}},\ \bibinfo
  {pages} {185301} (\bibinfo {year} {2022})}\BibitemShut {NoStop}%
\bibitem [{\citenamefont {Huang}\ \emph {et~al.}(2021)\citenamefont {Huang},
  \citenamefont {Luo}, \citenamefont {Wu},\ and\ \citenamefont
  {Xu}}]{Huang2021}%
  \BibitemOpen
  \bibfield  {author} {\bibinfo {author} {\bibfnamefont {G.-H.}\ \bibnamefont
  {Huang}}, \bibinfo {author} {\bibfnamefont {G.-Q.}\ \bibnamefont {Luo}},
  \bibinfo {author} {\bibfnamefont {Z.}~\bibnamefont {Wu}}, \ and\ \bibinfo
  {author} {\bibfnamefont {Z.-F.}\ \bibnamefont {Xu}},\ }\href {\doibase
  10.1103/PhysRevA.103.043328} {\bibfield  {journal} {\bibinfo  {journal}
  {Phys. Rev. A}\ }\textbf {\bibinfo {volume} {103}},\ \bibinfo {pages}
  {043328} (\bibinfo {year} {2021})}\BibitemShut {NoStop}%
\bibitem [{\citenamefont {Pan}\ \emph {et~al.}(2016)\citenamefont {Pan},
  \citenamefont {Zhang}, \citenamefont {Yi},\ and\ \citenamefont
  {Guo}}]{Pan2016}%
  \BibitemOpen
  \bibfield  {author} {\bibinfo {author} {\bibfnamefont {J.-S.}\ \bibnamefont
  {Pan}}, \bibinfo {author} {\bibfnamefont {W.}~\bibnamefont {Zhang}}, \bibinfo
  {author} {\bibfnamefont {W.}~\bibnamefont {Yi}}, \ and\ \bibinfo {author}
  {\bibfnamefont {G.-C.}\ \bibnamefont {Guo}},\ }\href {\doibase
  10.1103/PhysRevA.94.043619} {\bibfield  {journal} {\bibinfo  {journal} {Phys.
  Rev. A}\ }\textbf {\bibinfo {volume} {94}},\ \bibinfo {pages} {043619}
  (\bibinfo {year} {2016})}\BibitemShut {NoStop}%
\bibitem [{\citenamefont {Cornwell}(1984)}]{Group_Cornwell}%
  \BibitemOpen
  \bibfield  {author} {\bibinfo {author} {\bibfnamefont {J.~F.}\ \bibnamefont
  {Cornwell}},\ }\href@noop {} {\emph {\bibinfo {title} {Group Theory in
  Physics}}},\ Vol.~\bibinfo {volume} {1}\ (\bibinfo  {publisher} {Academic
  Press},\ \bibinfo {year} {1984})\BibitemShut {NoStop}%
\bibitem [{\citenamefont {Dresselhaus}\ \emph {et~al.}(2007)\citenamefont
  {Dresselhaus}, \citenamefont {Dresselhaus},\ and\ \citenamefont
  {Jorio}}]{Dresselhaus2007group}%
  \BibitemOpen
  \bibfield  {author} {\bibinfo {author} {\bibfnamefont {M.~S.}\ \bibnamefont
  {Dresselhaus}}, \bibinfo {author} {\bibfnamefont {G.}~\bibnamefont
  {Dresselhaus}}, \ and\ \bibinfo {author} {\bibfnamefont {A.}~\bibnamefont
  {Jorio}},\ }\href@noop {} {\emph {\bibinfo {title} {Group theory: application
  to the physics of condensed matter}}}\ (\bibinfo  {publisher} {Springer
  Science \& Business Media},\ \bibinfo {year} {2007})\BibitemShut {NoStop}%
\bibitem [{\citenamefont {Zhai}(2015)}]{Zhai_2015}%
  \BibitemOpen
  \bibfield  {author} {\bibinfo {author} {\bibfnamefont {H.}~\bibnamefont
  {Zhai}},\ }\href {\doibase 10.1088/0034-4885/78/2/026001} {\bibfield
  {journal} {\bibinfo  {journal} {Reports on Progress in Physics}\ }\textbf
  {\bibinfo {volume} {78}},\ \bibinfo {pages} {026001} (\bibinfo {year}
  {2015})}\BibitemShut {NoStop}%
\bibitem [{Note1()}]{Note1}%
  \BibitemOpen
  \bibinfo {note} {This phase transition can be viewed as a
  stripe-to-plane-wave phase transition with respect to the transformed
  Hamiltonian $\protect \hat H' = U\protect \hat H U^{-1}$.}\BibitemShut
  {Stop}%
\bibitem [{\citenamefont {Ikegami}\ \emph {et~al.}(2013)\citenamefont
  {Ikegami}, \citenamefont {Tsutsumi},\ and\ \citenamefont
  {Kono}}]{2013Ikegami}%
  \BibitemOpen
  \bibfield  {author} {\bibinfo {author} {\bibfnamefont {H.}~\bibnamefont
  {Ikegami}}, \bibinfo {author} {\bibfnamefont {Y.}~\bibnamefont {Tsutsumi}}, \
  and\ \bibinfo {author} {\bibfnamefont {K.}~\bibnamefont {Kono}},\ }\href
  {\doibase 10.1126/science.1236509} {\bibfield  {journal} {\bibinfo  {journal}
  {Science}\ }\textbf {\bibinfo {volume} {341}},\ \bibinfo {pages} {59}
  (\bibinfo {year} {2013})}\BibitemShut {NoStop}%
\bibitem [{\citenamefont {Kallin}\ and\ \citenamefont
  {Berlinsky}(2016)}]{2016Kallin}%
  \BibitemOpen
  \bibfield  {author} {\bibinfo {author} {\bibfnamefont {C.}~\bibnamefont
  {Kallin}}\ and\ \bibinfo {author} {\bibfnamefont {J.}~\bibnamefont
  {Berlinsky}},\ }\href {\doibase 10.1088/0034-4885/79/5/054502} {\bibfield
  {journal} {\bibinfo  {journal} {Reports on Progress in Physics}\ }\textbf
  {\bibinfo {volume} {79}},\ \bibinfo {pages} {054502} (\bibinfo {year}
  {2016})}\BibitemShut {NoStop}%
\bibitem [{\citenamefont {Tran}\ \emph {et~al.}(2017)\citenamefont {Tran},
  \citenamefont {Dauphin}, \citenamefont {Grushin}, \citenamefont {Zoller},\
  and\ \citenamefont {Goldman}}]{2017Tran}%
  \BibitemOpen
  \bibfield  {author} {\bibinfo {author} {\bibfnamefont {D.~T.}\ \bibnamefont
  {Tran}}, \bibinfo {author} {\bibfnamefont {A.}~\bibnamefont {Dauphin}},
  \bibinfo {author} {\bibfnamefont {A.~G.}\ \bibnamefont {Grushin}}, \bibinfo
  {author} {\bibfnamefont {P.}~\bibnamefont {Zoller}}, \ and\ \bibinfo {author}
  {\bibfnamefont {N.}~\bibnamefont {Goldman}},\ }\href {\doibase
  10.1126/sciadv.1701207} {\bibfield  {journal} {\bibinfo  {journal} {Science
  Advances}\ }\textbf {\bibinfo {volume} {3}},\ \bibinfo {pages} {1701207}
  (\bibinfo {year} {2017})}\BibitemShut {NoStop}%
\bibitem [{\citenamefont {Midtgaard}\ \emph {et~al.}(2020)\citenamefont
  {Midtgaard}, \citenamefont {Wu}, \citenamefont {Goldman},\ and\ \citenamefont
  {Bruun}}]{2020Midtgaard}%
  \BibitemOpen
  \bibfield  {author} {\bibinfo {author} {\bibfnamefont {J.~M.}\ \bibnamefont
  {Midtgaard}}, \bibinfo {author} {\bibfnamefont {Z.}~\bibnamefont {Wu}},
  \bibinfo {author} {\bibfnamefont {N.}~\bibnamefont {Goldman}}, \ and\
  \bibinfo {author} {\bibfnamefont {G.~M.}\ \bibnamefont {Bruun}},\ }\href
  {\doibase 10.1103/PhysRevResearch.2.033385} {\bibfield  {journal} {\bibinfo
  {journal} {Phys. Rev. Research}\ }\textbf {\bibinfo {volume} {2}},\ \bibinfo
  {pages} {033385} (\bibinfo {year} {2020})}\BibitemShut {NoStop}%
\end{thebibliography}
\end{document}